\documentclass[12pt,reqno]{article}

\usepackage{amsmath,amsthm,amssymb,epsfig,alltt}

\def\a{\alpha}
\def\b{\beta}
\def\d{{\delta}}

\def\e{\epsilon}
\def\p{\partial}
\def\m{\mu}
\def\n{\nu}
\def\t{\tau}
\def\th{\theta}
\def\s{\sigma}
\def\g{\gamma}

\def\r{\rho}

\def\G{{\Gamma}}
\def\half{\frac{1}{2}}

\def\tr{{\rm tr}}

\def\nn{\nonumber}

\def\2pap{2\pi\alpha^\prime}

\def\beq{\begin{eqnarray}}
 \def\eeq{\end{eqnarray}}
 \def\4pap{4\pi\a^\prime}

 \def\ap{{\a^\prime}}
 
 \def\zp{{z^\prime}}

 \def\bolk{{\boldsymbol k}}

 \def\bbP{{\boldsymbol P}}

 \def\bbV{{\boldsymbol V}}

\begin{document}


\title{Covariant open bosonic string field theory on multiple D-branes in the 
 proper-time gauge }

\author{Taejin Lee \\~~\\
Department of Physics, Kangwon National University, \\
Chuncheon 24341 Korea\\
email: taejin@kangwon.ac.kr}

\maketitle

\centerline{\bf Abstract}

We construct a covariant open bosonic string field theory on multiple D-branes, which reduces to a
non-Abelian group Yang-Mills gauge theory in the zero-slope limit. Making use of the first quantized open 
bosonic string in the proper time gauge, we convert the string amplitudes given by the Polyakov 
path integrals on string world sheets into those of the second quantized theory. 
The world sheet diagrams generated by the constructed open string field theory 
are planar in contrast to those of the Witten's cubic string field theory. However, the constructed 
string field theory is yet equivalent to the Witten's cubic string field theory. 
Having obtained planar diagrams, we may adopt the light-cone string field theory technique to 
calculate the multi-string scattering amplitudes with an arbitrary number of external 
strings. We examine in detail the three-string vertex diagram and the effective four-string vertex diagrams generated perturbatively by the three-string vertex at tree level. 
In the zero-slope limit, the string scattering amplitudes are identified precisely as those of 
non-Abelian Yang-Mills gauge theory if the external states are chosen to be massless vector particles. 

\vskip 2cm






\section{Introduction}

The non-Abelian group Yang-Mills gauge field theory \cite{Yang54} is the main pillar 
of the high energy physics. 
The string theory, which is the most promising framework of the unification of fundamental forces, should consistently reproduce the non-Abelian gauge field theory in the low energy limit. 
This condition is not an optional, but indispensable requirement, 
which must be satisfied by any open string theory. 
Taking this condition as a guiding principle, in the present work we construct a covariant interacting open string field theory on multiple D-branes \cite{Polchinski1995}. 
The first quantized string theory is formulated covariantly, as the string amplitude is defined as 
a path integral over the string world sheet with the covariant Polyakov string action \cite{Polyakov1981}. 
The covariant second quantized 
string theories, {\it i.e.}, the covariant string field theories, 
also have been proposed in the literature \cite{Siegel84a, Siegel84b, Witten1986, Hata1986, Witten92p}. However, 
it has not been fully established yet that the proposed covariant open string field theories yield the 
non-Abelian gauge field theories in the zero-slope limit which corresponds to the low energy limit in string theory. 

The non-Abelian local gauge invariance is often taken for granted in covariant string field theories. 
However, it is not a simple criterion that the string field theories may trivially meet. Unlike in the case of 
quantum field theories for point particles, which do not have internal structures, the local non-Abelian gauge
invariance is not manifest in the case of string field theories which deal with the extended objects, {\it i.e.}, strings. This is because that all higher mass level modes are not decoupled from the dynamics 
of low mass level modes even in the low energy region in the string theory. A series of studies on the 
low energy effective action of the open string field theory, based on the level truncation 
\cite{Kostecky90,Kostecky96,Sen99,Taylor2000} clearly reveal this distinctive feature of the string field theory.   

It is not difficult to construct covariant free open string field theories. 
The free string propagator can be 
expressed as a path integral over a surface of strip with two spatial boundaries. In the presence of 
multiple D-branes, two ends of the open string are attached on the D-branes and they 
necessarily carry the Chan-Paton factors \cite{chan-paton}. For $N$ D-branes, the open string fields take values in the Lie-algebra of $U(N)$ group. We choose the proper-time gauge condition to fix the reparameterization of the Polyakov action
on the string world sheet, which has been applied to the bosonic string \cite{Lee1988} 
and the super-string in the critical dimensions \cite{Lee1987} to construct the
Becchi-Rouet-Stora-Tyutin (BRST) \cite{Becchi1974, Becchi1975, Becchi1976,
Tyutin1975} invariant covariant free string field actions. If the 
Arnowitt-Deser-Misner (ADM) formulation \cite{Arnowitt1959} is adopted for the two-dimensional 
world sheet metric, the metric is parameterized by the lapse and shift functions. 
We can fix the reparameterization invariance in a covariant manner if we impose the 
the gauge fixing condition on the lapse and shift functions.  
In the proper-time gauge the lapse and shift functions are chosen to 
vanish except for the zero-mode of the lapse function. The world-sheet temporal coordinate 
becomes the proper-time \cite{Schwinger1951} in this gauge. The BRST ghost structure emerges 
as we evaluate the Faddeev-Popov determinant, which arises from fixing the reparameterization 
invariance by choosing the proper-time gauge condition. The advantage of the proper-time gauge is evident: 
The reparameterization degrees of freedom can be fixed completely in a covariant manner and the 
Polyakov string path integral over the strip can be directly converted into the free propagator, 
which leads us to a covariant, BRST invariant, free string field action of second quantized theory. 

In the proper-time gauge the lapse function, which is a constant, fixes the scale of energy along the 
direction of the proper-time while the scale of spatial 
coordinate on the world sheet, which corresponds to the length parameter of the conventional covariant 
string field theory, is inversely related to the scale of energy. Hence, we may be able to 
fix the length parameters on the world sheet completely by choosing the proper-time gauge. In the conventional
covariant string field theory of Hata, Itoh, Kugo, Kunitomo and Ogawa (HIKKO) \cite{Hata1986},
the length parameter has been introduced as an analogue of the light-cone momentum of 
the sting field theory in the light-cone gauge. 
Here we should note that the length parameter is not a free parameter in the proper-time gauge.
If the length parameters are introduced 
as free parameters of the theory, the string fields, consequently the massless states and other massive 
states of open string inevitably depend on the length parameters, which may be unphysical ones in the low energy sector. It is certainly an undesirable feature of the theory.  
The strings with different length parameters can be transformed onto each other by 
reparameterization. This implies that 
if we fix the reparameterization invariance by choosing the proper-time gauge, we may be able to
fix the length parameters in a way that is consistent and compatible with the covariance of the second 
quantized string theory. 
Because the length parameters are fixed in the proper-time gauge, the covariant open string field
theory to be proposed in the present paper is free of such unphysical parameters.

\begin{figure}[htbp]
   \begin {center}
    \epsfxsize=0.9\hsize
    
	\epsfbox{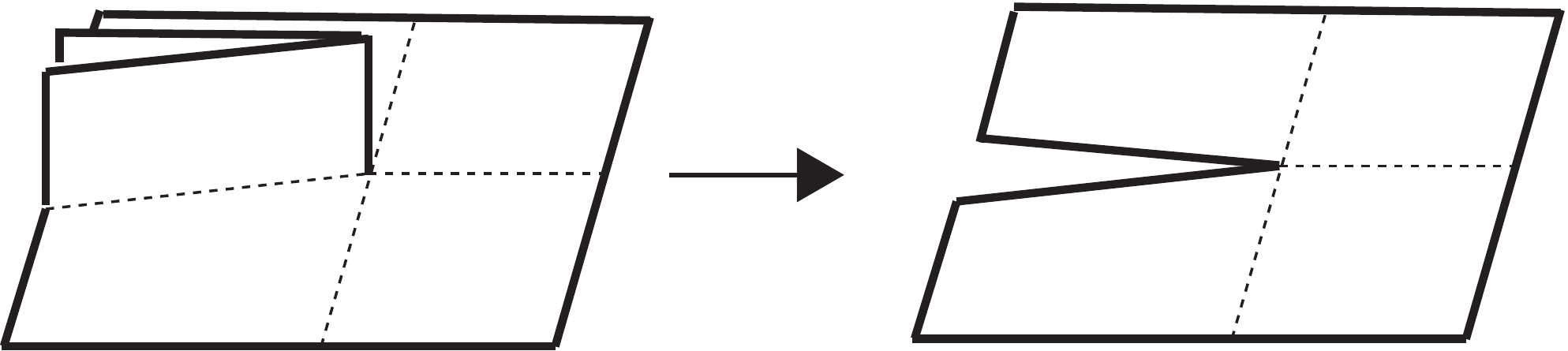}
   \end {center}
   \caption {\label{cubictoproper} Deformation of the cubic string vertex}
\end{figure}

The covariant cubic open string field theory proposed by Witten \cite{Witten1986} 
also does not contain the length parameters. This is because the common scale 
of energy and the length parameters of the cubic open string field theory are completely fixed.
In a recent work, we discussed a consistent deformation of the Witten's cubic open string field theory
\cite{Leedeform17} in such a way that the non-planar diagrams of multi-string scatterings of the cubic string 
field theory may be transformed into planar ones. For the three-string vertex and the four-string 
vertex we only need to alter the external string states by taking direct products with additional
Neumann states which do not carry momenta. Then, we may effectively remove 
some parts of the world sheets to 
transform the world sheet diagrams planar so that we may apply the light-cone string field theory technique 
\cite{Mandelstam1973, Mandelstam1974, Kaku1974a, Kaku1974b, Cremmer74, Cremmer75, Green1982, GreenSW87} to calculate the scattering amplitudes systematically. Once deforming the three-string 
scattering diagram of the cubic string field theory, we may identify it as that of the string field 
theory in the proper time gauge as depicted in Fig. \ref{cubictoproper}. Although the string field theory in the proper time gauge may be apparently equivalent to 
the covariantized light-cone string field theory of HIKKO with length parameters fixed, it is not 
necessary to take integration over the unphysical length parameters to make the theory 
BRST gauge invariant. To restore the BRST gauge invariant form, we only need to 
reattach the auxiliary patch \cite{Leedeform17} which does not contribute to scattering amplitude.

The world sheet geometry of the cubic string field theory is a cone with an excess angle. 
For the world sheet tree diagram with $N$ vertices, the excess angle is $N \pi$. 
Therefore, to obtain the scattering amplitudes of the cubic string field theory we have to deal with the 
Green's function on the non-planar space. Because the Green's functions on the non-planar space are not 
generally known, we may map the world sheet diagram onto planar spaces. Gross and Jevicki \cite{Grossjevicki87a, Grossjevicki87b} obtained the Fock space representation of the three-string vertex by mapping the world sheet diagram of six strings 
onto a unit disk. The Fock space representation of the four-string vertex has been found 
by Giddings \cite{Giddings86} by constructing a conformal mapping of the four-string diagram onto the 
upper half plane. To describe the non-planar spaces, they had to introduce additional 
structures: For the three-string vertex, an orbifold condtion was imposed and for the four-string 
vertex, branch cuts were introduced. However, it is not known how to extend these procedures for more 
general string scattering diagrams with an arbitrary number of external strings.
Because there is no known relationship between two conformal mappings, it is also difficult to fix the relative strength between three-string scattering amplitude and the four-string scattering amplitude.  
Thus, we cannot confirm that the non-Abelian local gauge invariance emerges in the low energy limit of the 
cubic string field theory. It is certainly desirable to develop a new systematic approach. 
In view of this we propose the open string field theory in the proper time gauge
as a practical tool to study the perturbation theory of covariant interacting strings. 

As a consistency check, we show in the Ref. \cite{Leedeform17} that the open string field theory 
in the proper-time gauge yields the non-Abelian Yang-Mills action in the zero-slope limit. 
Here, we present details of the reduction: The first step is to identify the 
three-gauge-field vertex of the Yang-Mills theory by taking the zero-slope limit 
of the three-string vertex in the proper time gauge with the correct relationship between 
the Yang-Mills coupling and the string interaction coupling. The second step, which is non-trivial, is to confirm whether the covariant string field theory reproduces also the contact four-gauge-field-interaction term of the Yang-Mills theory in the zero-slope limit. Because the open string is an extended object, 
in the zero-slope limit the four-string scattering amplitudes, which the three-string vertex generates,
may contain not only that of the contact four-gauge-field vertex but also those diagrams, which consist
of two three-gauge-field vertices and an intermediate gauge field propagator. 
We shall also present explicit expressions of the Neumann functions of the three-string vertex and the four-string vertex which would be useful to study the various particle scattering amplitudes as applications of 
the string field theory in the proper time gauge.

\section{Open String in the Proper-Time Gauge}

The covariant first quantized string theory is founded on the covariant geometrical Polyakov action.
The string amplitudes are defined in a covariant manner by using the path integrals over the two-dimensional
Riemann surfaces, which correspond to the string world sheets. On the other hand, the covariant second quantized 
theory, {\it i.e.}, the string field theory is built on the BRST (Becchi-Rouet-Stora-Tyutin) formalism. 
In the previous works \cite{Lee1988,Lee1987}, we established a direct connection between the geometrical approach of the first quantized theory based on the Polyakov path integral and the second quantized theory based on the BRST symmetry, towards the covariant string theories in the cases of the free string theories. In the present work
we shall extend the previous works to the interacting bosonic string on multiple D-branes. In this respect, it may be appropriate to begin with the Brink-Di Vecchia-Howe-Deser-Zumino action or Polyakov action 
of bosonic string 
\beq
S = -\frac{1}{4\pi \ap} \int_M d\t d\s \sqrt{-h} h^{\a\b} \frac{\p X^I}{\p \s^\a} \frac{\p X^J}{\p \s^\b} \eta_{IJ}, ~~~~~ I, J = 0, \dots , d-1 
\eeq
where $\a^\prime$ is the Regge slope parameter and $\s^1 = \t$,  $\s^2 = \s$. Here $d=26$ for the 
open bosonic string and $d=10$ for the open super-string.   
We use the space-time metric $\eta_{IJ} = {\rm diag}(-,+,+, \cdots , +)$. If the open string is defined on $Dp$-branes, we impose the Neumann boundary conditions on the string coordinates, tangential to the D-branes and Dirichlet boundary conditions on the string coordinates, normal to the D-branes 
\begin{subequations}
\beq
\frac{\p X^\m}{\p \s} \Bigl\vert_{\s=0, \pi} &=& 0, ~~~\text{for} ~~\mu = 0, 1, \dots, p, \\
X^a\Bigl\vert_{\s=0, \pi} &=& \bar x^a, ~~~\text{for} ~~a = p+1, \dots, d-1.
\eeq
\end{subequations}
For the sake of simplicity, in this work we shall consider the space filling D-brane only, 
because our main concern is to construct a covariant open bosonic string field theory, which reduces to the non-Abelian group Yang-Mills gauge field theory
\beq
S = -\frac{1}{4\pi \ap} \int_M d\t d\s \sqrt{-h} h^{\a\b} \frac{\p X^\m}{\p \s^\a} \frac{\p X^\n}{\p \s^\b} \eta_{\m\n}, ~~~~~ \m, \n = 0, \dots , d-1 . 
\eeq

Applying the ADM formulation, we may write the two-dimensional world sheet metric in terms of the lapse and shift functions as follows
\beq
\sqrt{-h} h^{\a\b} = \frac{1}{N_1} \begin{pmatrix} -1 & N_2 \\ N_2 & (N_1)^2 - (N_2)^2 \end{pmatrix}.
\eeq
Subsequently, the classical Polyakov Lagrangian density is read as 
\beq
{\cal L} = \frac{1}{2N_1} \left[\p_\t X^\m \p_\t X_\m -2 N_2 \p_\t X^\m \p_\s X_\m -
\left((N_1)^2 -(N_2)^2 \right) \p_\s X^{\m} \p_\s X_\m \right].
\eeq
We choose the units in which the string tension $T = \frac{1}{2\pi \ap} = 1$ for convenience.
Defining the canonical momenta
\beq
P^\m =  \frac{1}{N_1} \left(\p_\t X^\m - N_2 \p_\s X^\m \right),
\eeq
we obtain the canonical Hamiltonian density
\beq
{H} &=&  P^\m \p_\t X_\m - {\cal L} \nn\\
&=&  \frac{N_1}{2} \left[P^\m P_\m +\p_\s X^{\m} \p_\s X_\m \right]
+ N_2 P^\m \p_\s X_\m . 
\eeq
The path integral over a strip $M$, representing the free propagator of the open string,
is given as
\begin{subequations} 
\beq
G[X^\m_1;X^\m_2] &=& \int D[N] D[X] \exp \left[iS \right] \nn\\
&=& \int D[N] D[P,X] \exp \left[i \int_M d\t d\s \left(P^\m \dot{X_\m} - N^1 \varphi_1 - N^2 \varphi_2\right) \right], \label{strip}\\
\varphi_1 &=& \half \left(P^\m P_\m + \p_\s X^\m \p_\s X_\m \right),\label{proper1}\\
\varphi_2 &=& P^\m \p_\s X_\m . \label{proper2}
\eeq
\end{subequations}
At two ends of the strip, the string coordinates take the values of $X^\m_1$ and $X^\m_2$
respectively. 
The lapse and shift functions turn out to be the Lagrangian multipliers of the canonical quantization. 

The two first-class constraints $\varphi^1$ and $\varphi^2$ generate the reparameterization under which 
the lapse and shift functions transform as 
\begin{subequations}
\beq
\d N_1 &=& \p_\t \e_1 + \p_\s N_1 \e_2 - N_1 \p_\s \e_2 + \p_\s N_2  \e_1 - N_2 \p_\s \e_1,  \\
\d N_2 &=& \p_\t \e_2 + \p_\s N_1 \e_1 - N_1 \p_\s \e_1 + \p_\s N_2 \e_2 - N_2 \p_\s \e_2 ,
\eeq
\end{subequations}
where $\e_1 \vert_{\p M} = 0$. The proper-time gauge condition is the gauge fixing condition,
which is imposed on the lapse and shift functions 
\beq \label{propercondition}
\p_\t N_{10} = 0, ~~N_{1n} = 0, ~~ N_{2n}= 0, ~~ n \not=0,
\eeq
where $N_\a = \sum_n N_{\a n} e^{in\s}$, $\a=1, 2$. ($N_{20} = 0$ for open string.)
In the proper-time gauge the two-dimensional world sheet metric is fixed by a constant $n_1$, (the zero-mode 
of the lapse function) as 
\beq
\sqrt{-h} h^{\a\b} = \frac{1}{n_1} \begin{pmatrix} -1 & 0 \\ 0 & n^2_1  
\end{pmatrix}.
\eeq
Using the conformal invariance at the critical dimensions, we also fix the conformal factor
by imposing a condition such that $\sqrt{-h} = 1$.  It leads us to the following two-dimensional line element
\begin{subequations} 
\beq
(dl)^2 &=& h_{\a\b} d\s^\a d\s^\b = -n_1 d\t d\t + \frac{1}{n_1} d\s d\s, \\
h_{\a\b} &=& \begin{pmatrix} -n_1 & 0 \\
0 & 1/n_1 \end{pmatrix}. 
\eeq
\end{subequations}
On the two-dimensional world sheet, we find that in the proper-time gauge, the scale in the temporal ($\t$) direction is 
inversely related to the scale factor in the spatial ($\s$) direction.

The proper-time gauge fixing condition given by Eqs.~(\ref{propercondition}) 
fixes the degrees of freedom of 
reparameterization consistently and completely \cite{Lee1988}. The gauge fixing by using the proper-time gauge condition results in the following Faddeev-Popov determinant in the measure,
\beq
\det \left[ \frac{\d N_\a}{\d \e_\b} \right] &=& \det\left[\p^2_\t\right] \prod_{n=1} \det 
\left[\begin{array}{cc} \p_\t & -in n_1 \\ -in n_1 & \p_\t \end{array}\right] \nn\\
&=& T \prod_{n=1} \det \left[\begin{array}{cc} \p_\t + in n_1 & 0 \\ 0 &  \p_\t -in n_1  \end{array}\right] 
\eeq
where $T= |\t_2 - \t_1|$.
The action of the BRST ghost fields arises from representing the Faddeev-Popov determinant by a path integral,
\begin{subequations} 
\beq
\det \left[ \frac{\d N_\a}{\d \e_\b} \right] &=& T \int D[c, d ] 
\exp \left[\int^{\t_2}_{\t_1}  d\t\sum_{n=1} \left( c_n \dot{d}_n  
-in_1 L^{gh}_0 \right) \right], \\
L^{gh}_0 &=& \sum_{n=1} n\, c_n d_n ,
\eeq
\end{subequations}
where $c_n$ and $d_n$ are anti-commuting BRST ghost fields. 
Evaluating the path integral over a strip Eq. (\ref{strip}) in the proper-time gauge explicitly brings us 
to the open string propagator of the second quantized theory
\begin{subequations} 
\beq
G[X^\m_1, c_1, d_1 ; X^\n_2, c_2, d_2] &=& \int^\infty_0 ds \int D[X, P] D[c, \bar c, d, \bar d] \nn\\
&& \exp \Biggl[ i \int^s_0 ds^\prime \left(\sum_{n=1} \left(P_n \cdot \dot{X}_n - i 
c_n \dot{\bar c}_n -i \bar d_n \dot{d}_n \right)\right) -L_0 - L^{gh}_0 \Biggr] \nn\\
&=& \int^\infty_0 ds \langle X^\m_1, c_i, d_1 \vert \exp \left[-is \left(L_0 + L_{gh} -i\e \right) \right]
\vert X^\n_2, c_2, d_2 \rangle \nn\\
&=&  \langle X^\m_1, c_1, d_1 \vert \frac{1}{L_0 + L^{gh}_0 -i\e}\vert X^\n_2, c_2, d_2 \rangle,  \\
L_0 &=& \sum_{n=1} \left(P^2_n + n^2 X^2_n \right),
\eeq
\end{subequations}
where $s = T n_1$ is the proper-time. 
\begin{figure}[htbp]
   \begin {center}
    \epsfxsize=0.5\hsize

	\epsfbox{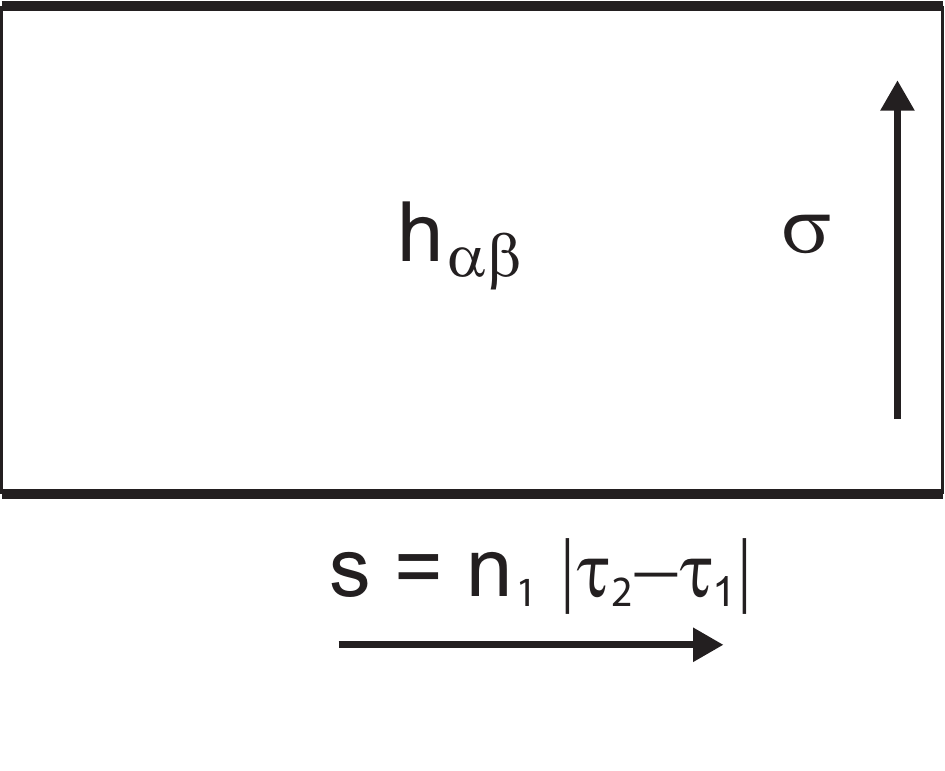}
   \end {center}
   \caption {\label{open1} String propagator}
\end{figure}
It is clear that the propagator 
$G[X^\m_1, c_1, d_1 ; X^\n_2, c_2, d_2]$ 
can be identified as the string propagator of the second quantized theory if we rewrite it as follows
\beq
G[X^\m_1, c_1, d_1 ; X^\n_2, c_2, d_2] &=& i \langle {\cal T} \Psi[X_2, c_2, d_2] \Psi[X_1, c_1 , d_1 ] \rangle \nn\\ &=& i \int D[\Psi] \,\, \Psi[X_2, c_2, d_2] \Psi[X_1, c_1, d_1 ] \nn\\
&&
\exp \left\{ - \frac{i}{2} \int D[X, c, d] \Psi\left(L_0+L^{gh}_0 \right) \Psi \right\}. 
\eeq
We note that the string propagator and the free string field action do not depend on the lapse function 
constant $n_1$ explicitly. 

Now we introduce the non-Abelian group indices for the string fields. On the multiple D-branes
the string field $\Psi$ carries the group indices
\beq
\Psi[X]  =\frac{1}{\sqrt{2}} \Psi^0 [X] + \Psi^a[X] T^a,  ~~~ a =1, \dots, N^2-1 ,
\eeq
where $\Psi^0$ is the $U(1)$ component and $\Psi^a$, $a =1, \dots, N^2-1$ are the $SU(N)$ components. 
Because the end point of open string is attached on one of the $N$ D-branes, 
the open string has $N^2$ different quantum states. 
We choose the adjoint representation to describe the quantum state of the open string field
appropriately. 
For the $SU(N)$ group 
\begin{subequations}
\beq
(T^a)_{bc} &=& -i f^{abc}, ~~~ \tr (T^a T^b) = \half \d_{ab}, \\
T^a T^b &=& \frac{1}{2} \left[T^a, T^b \right]+ \frac{1}{2} \left\{ T^a, T^b \right\}, \\
\left[T^a, T^b \right]&=& i \sum_{c} f^{abc} T^c . 
\eeq
\end{subequations}
Introducing the group indices, we may rewrite the string propagator as 
\beq
G^{ab}[X_1,\xi_1; X_2, \xi_2] &=&  i \langle {\cal T} \Psi^{a}[X_2, \xi_2] \Psi^b[X_1, \xi_1 ] \rangle \nn\\ &=& i \int D[\Psi] \,\, \Psi^{a}[X_2, \xi_2] \Psi^b[X_1, \xi_1 ] 
\exp \Bigl\{ -i {\cal S}_0 \Bigr\}
\eeq
where we denote the BRST ghost fields collectively by $\xi$. 
In accordance with the string propagator, which carries the group indices, 
the string free field action may be written as 
\beq
{\cal S}_0 &=& \int D[X,\xi] \tr \left(\Psi \left(L_0+L^{gh}_0 \right) \Psi\right), \nn\\
&=& \int D[X,\xi]\,\, \frac{1}{2} \left\{ \Psi^{0}  \left(L_0+L^{gh}_0 \right) \Psi^0 + 
\sum_{a=1}^{N^2-1} \Psi^{a}  \left(L_0+L^{gh}_0 \right) \Psi^a \right\}.
\eeq


\section{Interacting Open String in the Proper Time Gauge}
       
On the two-dimensional world sheet the line element along the spatial direction
is given by 
\beq
\Delta = \frac{1}{\sqrt{n_1}} |\s^\prime - \s| .
\eeq
The basic interactions of open string theory are processes of joining and splitting by which two open 
strings join together to make one or one string splits into two.  
Both processes are described by the three-string vertex. (See Fig. 2.)
\begin{figure}[htbp]
   \begin {center}
    \epsfxsize=0.5\hsize

	\epsfbox{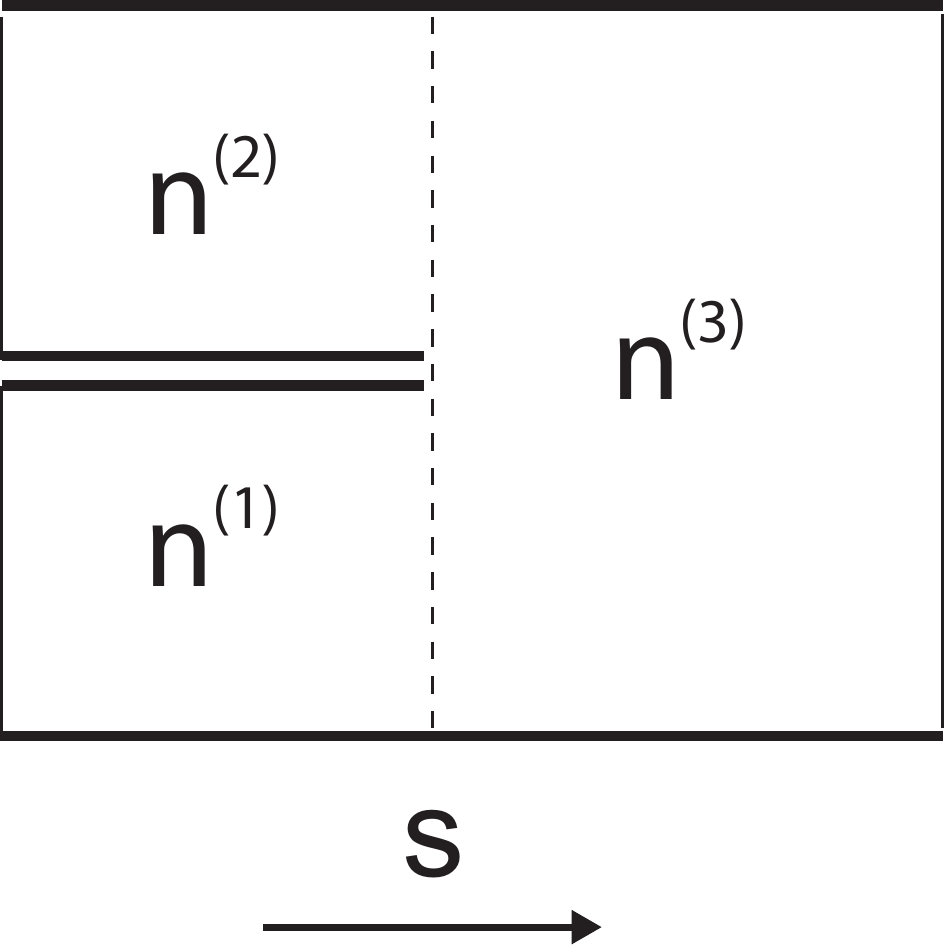}
   \end {center}
   \caption {\label{open2} Cubic string vertex in the proper time gauge}
\end{figure}
We may choose  a common proper-time 
for all three strings on the world sheet of three-string vertex, so that the energy scale would be same
at the junction of three strings
\beq
n^{(1)}_1= n_1^{(2)} =n_1^{(3)}=n_1.
\eeq 
Because the sum of the actual lengths of the first two open strings must be 
that of the third string at the junction of three strings,
\beq
\frac{\pi}{\sqrt{n_1}}  + \frac{\pi}{\sqrt{n_1}} = \frac{2\pi}{\sqrt{n_1}} . 
\eeq
Thus, we may choose the ranges of the spatial coordinates on the world sheet patches of individual 
strings as follows
\beq
\s_1 \in [0,\pi],~~ \s_2 \in [0,\pi], ~~\s_3 \in [0,2\pi].
\eeq
It is equivalent to fixing the length parameters of the HIKKO covariant open string field theory as 
\beq
\a_1 = \a_2 =1, ~~ \a_3 =-2.
\eeq
The spatial coordinates $\s_i$, $i=1, 2, 3$ on the three patches of the string world sheet are 
related to each other as 
\beq
\s = \begin{cases}  \s_1, \quad & \text{for} ~~ 0 \le \s \le \pi \\
\s_2 + \pi, \quad & \text{for} ~~ \pi \le \s \le 2\pi \\
2\pi -\s_3, \quad & \text{for}~~ 0\le \s \le 2\pi .  \end{cases} 
\eeq
(See Fig. 3.)
\begin{figure}[htbp]
   \begin {center}
    \epsfxsize=0.5\hsize

	\epsfbox{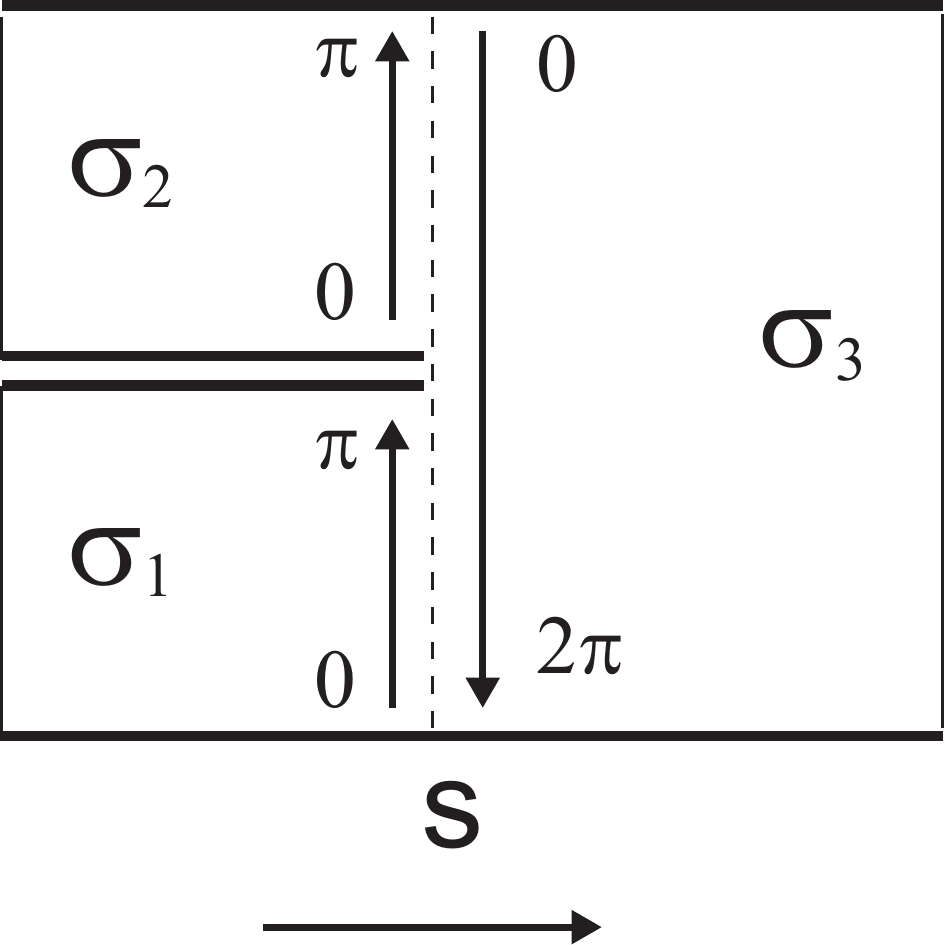}
   \end {center}
   \caption {\label{open3} Spatial coordinates on the world sheet of the three-string vertex}
\end{figure}

\subsection{Three-string vertex in the proper-time gauge}

As in the case of the free string field action, we shall abbreviate the group indices of the string fields first and restore them later after the three-string overlapping function on the string configuration space is 
completely fixed. 
The three-string interaction term may be written in the proper-time gauge as
\begin{subequations} 
\beq
\bbV_{[3]} &=& \int \prod_{r=1}^3 D[X^{(r)}] \delta[X^{(1)},X^{(2)};X^{(3)}]
\Psi[X^{(1)}]\Psi[X^{(2)}]\Psi^3[X^{(3)}], \\
\delta[X^{(1)},X^{(2)};X^{(3)}] &=& \prod_\s \delta \left[
X^{(3)}(\s_3)\Theta_3 - X^{(1)}(\s_1) \Theta_1- X^{(2)}(\s_2) \Theta_2\right],
\eeq
\end{subequations}
where
\beq
\Theta_1 = \th(\pi  - \s), \quad \Theta_2 = \th(\s-\pi), \quad \Theta_3 = \Theta_1 + \Theta_2 = 1 .
\eeq
Because we only concern reduction of the open string field theory to the 
non-Abelian group Yang-Mills field theory in the zero-slope limit, 
we will treat the string fields as functionals of
the string coordinates $X^{(r)\m}$, $r=1,2,3$ only, suppressing the ghost coordinates $\xi^{(r)}$. 
We define the open string coordinates and momentum variables in terms of the 
normal mode oscillators on each world sheet patch,  
$\s_r \in [0,\pi |\a_r|]$, $r=1, 2, 3$, which satisfy the Neumann boundary condition as
\begin{subequations}
\beq
X^{(r)}(\s_r) 
&=& x^{(r)} + 2\sum_{n=1} \frac{1}{\sqrt{n}} x^{(r)}_n  \cos \left(n \frac{\s_r}{|\a_r|}\right) ,\nn\\
&=& x^{(r)} + \sum_{n=1} \frac{i}{\sqrt{n}} \left(a^{(r)}_n - a^{(r)\dag}_n \right) 
\cos \left(n \frac{\s_r}{|\a_r|}\right), \\
P^{(r)}(\s_r) 
&=& \frac{1}{\pi|\a_r|} \left(p^{(r)} + 
\sum_{n=1} \sqrt{n} \,p^{(r)}_n \cos \left(n \frac{\s_r}{|\a_r|}\right)\right) \nn\\
&=&\frac{1}{\pi |\a_r|} \left(p^{(r)}+ \sum_{n=1} \sqrt{n} \left(a^{(r)}_n + a^{(r)\dag}_n \right) 
\cos \left(n \frac{\s_r}{|\a_r|}\right) \right),
\eeq
\end{subequations}
where $r =1, 2, 3$ and $\a_1=1$, $\a_2=1$, $\a_3=-2$. 

The overlapping function in the coordinate space, $\delta[X^{(1)}_1,X^{(2)}_2;X^{(3)}_3]$
may be written in momentum space as follows \cite{Green1982} 
\beq
\delta[P^{(1)},P^{(2)};P^{(3)}] &=& \int \prod_{r=1}^3 D[X^{(r)}] \delta[X^{(1)},X^{(2)};X^{(3)}] \nn\\
&&~~~~~\exp\left\{i \int^{2\pi}_0 d\s
\sum_{r=1}^3 P^{(r)}(\s) \cdot X^{(r)}(\s) \Theta_r \right\} \nn\\
&=& \prod_{\s} \delta\Bigl[P^{(3)}(\s_3)\Theta_3 + P^{(1)}(\s_1)\Theta_1 +
P^{(2)}(\s_2)\Theta_2\Bigr] \nn\\
&=& \prod_{0\le\s \le\pi} \delta\left(P^{(1)}(\s_1)+P^{(3)}(\s_3)\right) 
\prod_{\pi \le \s \le 2\pi}\delta\left(P^{(2)}(\s_2)+P^{(3)}(\s_3)\right).
\eeq
Making use of the normal mode expansions of the string momenta, $P^{(r)}$, $r=1, 2, 3$, 
\begin{subequations}
\beq
p^{(r)} &=& \int^{\pi |\a_r|}_0 d\s_r P^{(r)}(\s_r), \\
p^{(r)}_m &=& \frac{2}{\sqrt{m}}\int^{\pi |\a_r|}_0 d\s_r P^{(r)} (\s_r) \cos \frac{m\s_r}{|\a_r|},
\eeq
\end{subequations}
we may rewrite the overlapping condition in the momentum space as \cite{Green1982}
\begin{subequations} 
\beq
p^{(1)}+p^{(2)}+ p^{(3)} &=& 0, \\
p^{(3)}_m + \sum_{n=1}^\infty \left(A^{(1)}_{mn} p^{(1)}_n + A^{(2)}_{mn} p^{(2)}_n \right)+
B^{(1)}_m p^{(1)} + B^{(2)}_m p^{(2)} &=& 0 ,~~~m \not=0 \label{momentumcondition}
\eeq
\end{subequations}
where in the proper-time gauge for $n \not= \frac{m}{2}$
\begin{subequations}
\beq
A^{(1)}_{mn} &=& - \frac{(-1)^{m+n}}{\pi} \sqrt{mn}\frac{\sin \frac{m\pi}{2}}{n^2-\frac{m^2}{4}} , \\
A^{(2)}_{mn} &=& \frac{(-1)^{m}}{\pi} \sqrt{mn}  \frac{\sin \frac{m\pi}{2}}{n^2-\frac{m^2}{4}},\\
B^{(1)}_m &=& \frac{(-1)^m}{\pi} \frac{4}{m\sqrt{m}} \sin \frac{m\pi}{2}, \\
B^{(2)}_m &=& -\frac{(-1)^m}{\pi} \frac{4}{m\sqrt{m}} \sin \frac{m\pi}{2},
\eeq
and for $m=2n$ 
\beq
A^{(1)}_{2n,n} &=& \frac{1}{\sqrt{2}}, ~~~
A^{(2)}_{2n,n} = \frac{(-1)^n}{\sqrt{2}}.
\eeq
\end{subequations}
The zero-mode part of Eq. (\ref{momentumcondition}) is given by
\begin{subequations}
\beq
p^{(1)}B^{(1)}_m + p^{(2)} B^{(2)}_m 
= \bbP B_m, \\
\bbP = p^{(2)} -p^{(1)}, ~~~ B_m = - \frac{4}{\pi} \frac{(-1)^m}{m\sqrt{m}} \sin \frac{m\pi}{2} . 
\eeq
\end{subequations}

It is more convenient to represent the three-string vertex in the Fock space. 
Using the momentum-space harmonic oscillator wave function given as 
\begin{subequations}
\beq
\psi_n(p) &=& \langle n \vert p \rangle, \\
\vert p \rangle &=& (\text{constant}) \exp \left(- \frac{1}{4} p^2 + p \, a^\dag - \half a^\dag a^\dag \right) \vert 0 \rangle , 
\eeq
\end{subequations}
we are able to construct the Fock-space representation of the vertex operator $\bbV_{[3]}$. 
If we expand the momentum eigenstate in the basis of the number operator eigenstates, we have 
\beq
\vert p \rangle = \sum_n \vert n \rangle \langle n \vert p \rangle = \sum_n \vert n \rangle \psi_n(p) . \label{momentumeigen}
\eeq
The Fock space representation of the three-string vertex $\bbV_{[3]}$ follows from 
the momentum space representation of the overlapping condition Eq. (\ref{momentumcondition}) 
and the 
momentum representation of the harmonic oscillator Eq. (\ref{momentumeigen}), 
\beq
\vert \bbV_{[3]} \rangle &=&\int \prod_r D[P^{(r)}] (2\pi)^d \d \left( \sum_{r=1}^3 p_r \right) \nn\\
&& ~~~\exp \left[ 
\sum_{n=1}^\infty \sum_{r=1}^3 \left(- \frac{1}{4} \left(p^{(r)}_n \right)^2 +p^{(r)}_n a^{(r)\dag}_n - 
\half \left(a^{(r)\dag}_n\right)^2 \right)\right] \vert 0 \rangle \nn\\
&&~~~ \prod_{m=1} \delta\left[p^{(3)}_m + \sum_{n=1}^\infty \left(A^{(1)}_{mn} p^{(1)}_n + A^{(2)}_{mn} p^{(2)}_n \right)+B^{(1)}_m p^{(1)} + B^{(2)}_m p^{(2)} \right].
\eeq
Performing the Gaussian integrations over the momentum modes, we may obtain the Fock space representation of $\vert \bbV_{[3]} \rangle$, in the proper-time gauge. However, this task is rather involved. The most efficient way is to map the world sheet into the upper half complex plane where the Neumann function 
takes a simple form. 

\begin{figure}[htbp]
   \begin {center}
    \epsfxsize=0.9\hsize

	\epsfbox{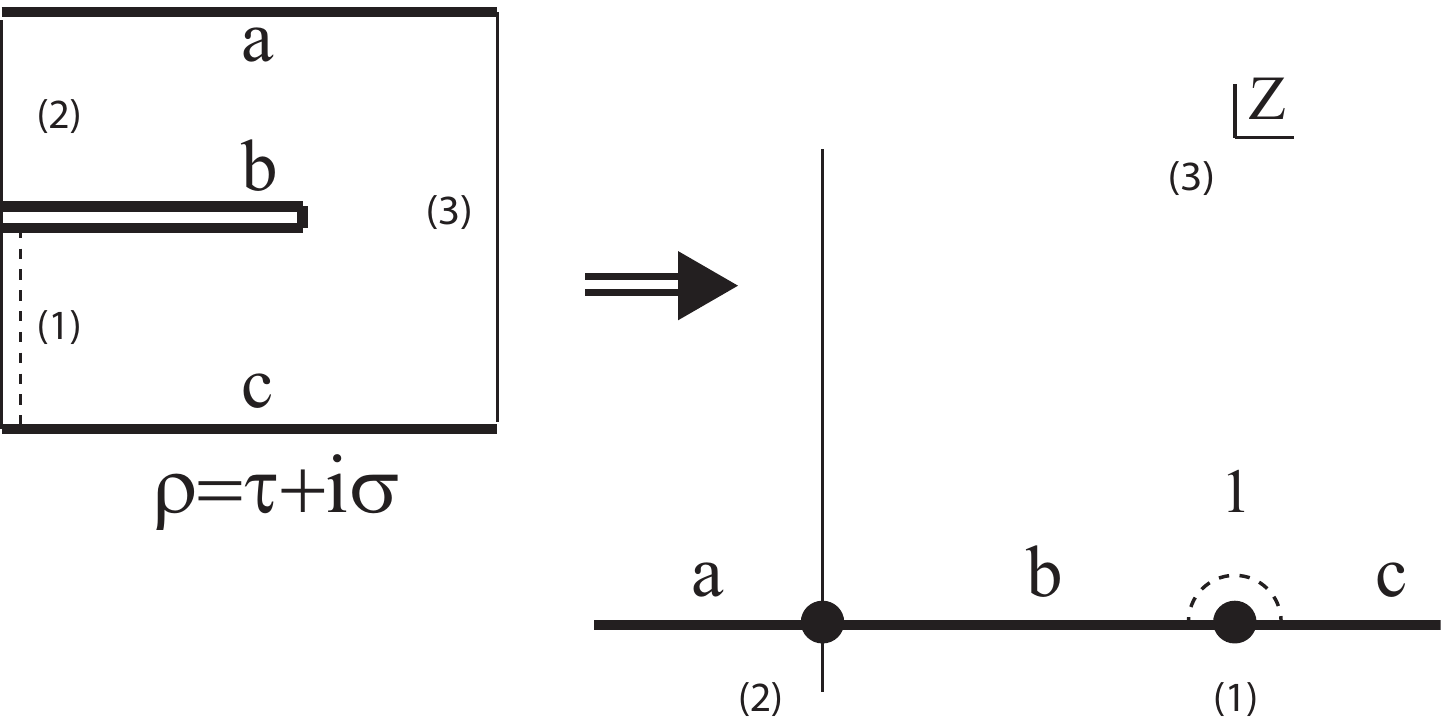}
   \end {center}
   \caption {\label{neumann3} Mapping of the three-string vertex diagram onto the upper half complex plane}
\end{figure}

We may map the planar world sheet onto the upper half complex plane by using the 
Schwarz-Christoffel trnasformation \cite{Mandelstam1973, Cremmer74, Cremmer75} given as 
\beq
\rho = \t + i\s = \sum_r \a_r \ln (z- Z_r) 
\eeq 
where $\rho$ is a global coordinate defined on the planar diagram. 
The boundaries which consists of the world lines of the end-points of the three open strings are mapped 
to form the real line and the three open strings in the asympotic regions are mapped onto small semi-circles around the points $z= Z_r$, $r =1, 2, 3$ on the real line. For convenience we choose 
\beq
Z_1 =1, ~~ Z_2 = 0, ~~ Z_3 = \infty .
\eeq 
For the three-string vertex in the proper-time gauge, 
\beq
\rho = \ln (z-1) + \ln z .  
\eeq 
The interaction point (the midpoint at the interaction time) is determined by 
\beq
\frac{\p\rho}{\p z} = 0. 
\eeq 
Its solution is found as $z_0 = 1/2$. Accordingly, the interaction time is fixed as 
\beq
\t_0 = \text{Re}\, \rho_0 = \sum_{r=1}^3 \a_r \ln \a_r = -2\ln 2. 
\eeq 
The local coordinates $\zeta_r = \xi_r + i \eta_r$, $r=1, 2, 3$ 
defined on invidual string world sheet patches are related to $z$ as follows: 
\begin{subequations}
\beq
e^{-\zeta_1} &=& e^{\t_0} \frac{1}{z(z-1)}, \\
e^{-\zeta_2} &=& - e^{\t_0} \frac{1}{z(z-1)} \\
e^{-\zeta_3} &=& - e^{- \frac{\t_0}{2}} \sqrt{z(z-1)} 
\eeq
\end{subequations}   
\begin{figure}[htbp]
   \begin {center}
    \epsfxsize=0.6\hsize

	\epsfbox{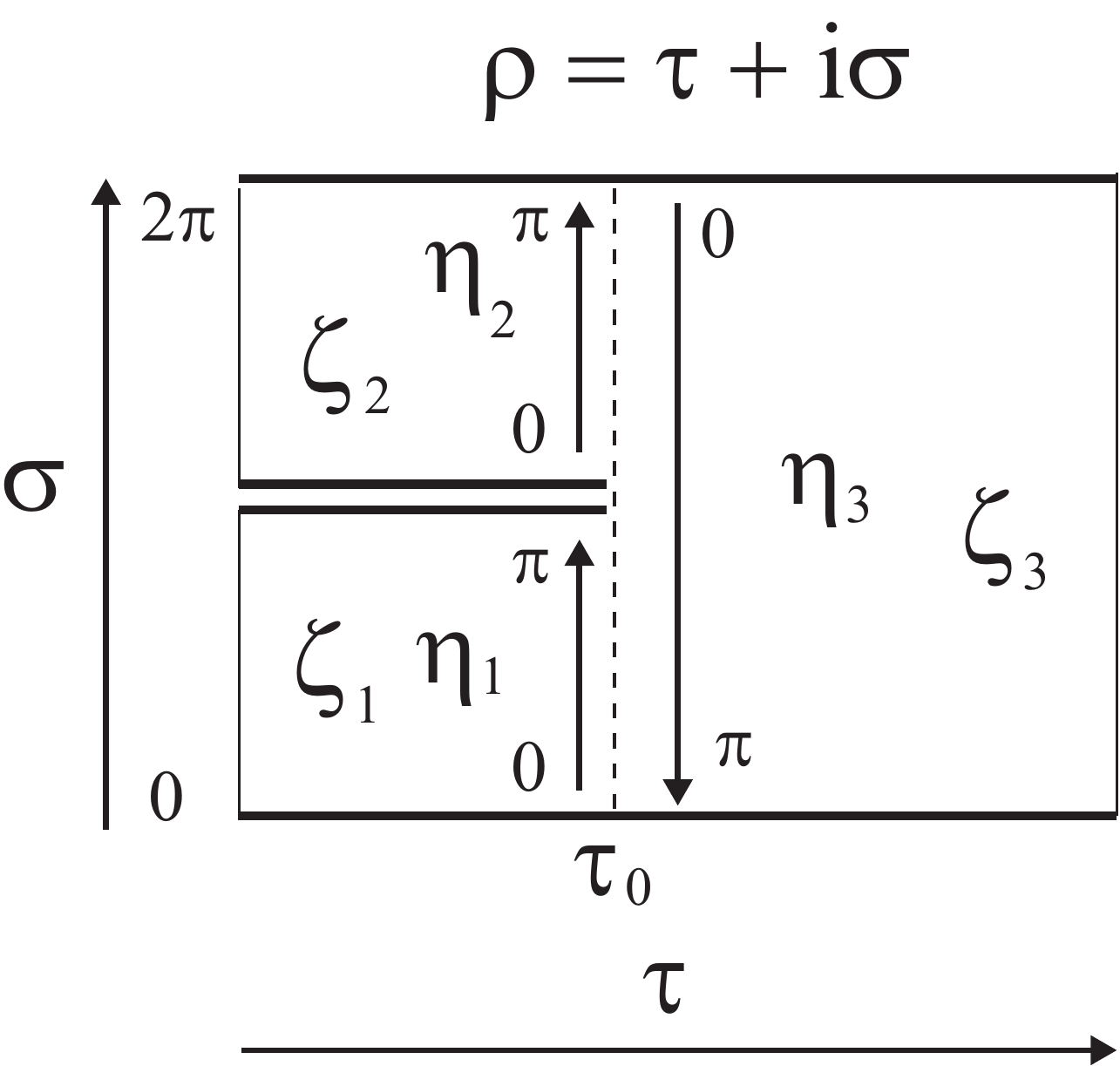}
   \end {center}
   \caption {\label{3vertex} Local coordinates on the planar diagram}
\end{figure}

The Neumann functions are defined as Fourier components of the Green's function with the 
Neumann boundary condition on the upper half complex plane \cite{Mandelstam1973, Cremmer74, GreenSW87}
\beq \label{neumandef}
N(\r_r, \rho^\prime_s) &=& \ln \vert z - \zp \vert + \ln \vert z - \zp^* \vert \nn\\
&=& - \d_{rs} \left\{ \sum_{n \ge 1} \frac{2}{n} e^{-n\vert \xi_r - \xi^\prime_s \vert} \cos (n\eta_r) 
\cos (n \eta^\prime_s) - 2 \text{max} \left(\xi_r, \xi^\prime_s \right) \right\} \nn\\
&& + 2 \sum_{n, m \ge 0} \bar N^{rs}_{nm} e^{n \xi_r + m \xi^\prime_s} \cos (n\eta_r) \cos (m\eta^\prime_s)
\eeq
where $\rho_r$ and $\rho^\prime_s$ lie in the regions of the $r$-th and $s$-th string patches respectively.
Defining 
\begin{subequations}
\beq
\bar N^r_n &=& \frac{1}{\a_r} f_n \left(-\frac{\a_{r+1}}{\a_r} \right) e^{\frac{n\t_0}{\a_r}}, ~~~ 
\a_1 = \a_4 = 1, ~ \a_2 =1, ~ \a_3 = -2,  \\
f_n(\g) &=& \frac{1}{n!} (n\g -1) (n\g -2) \cdots (n\g -n +1) \nn\\
&=& \frac{1}{n!} \frac{\G(n\g)}{\G(n\g +1 -n)} = \frac{(-1)^{n-1}}{n!} \frac{\G(n-n\g)}{\G(1-n\g)},
\eeq
\end{subequations}
we may write the Neumann functions for the three-string vertex \cite{Mandelstam1973, Cremmer74, GreenSW87} in the proper time gauge as 
\begin{subequations}
\beq
\bar N^{rs}_{nm} &=& 2 \left( \frac{\a_r}{n} + \frac{\a_s}{m} \right)^{-1} \bar N^r_n \bar N^s_m,
~~~ n, m \ge 1 , \\
\bar N^{r1}_{n0} &=& - \bar N^r_n, ~~ \bar N^{r2}_{n0} = \bar N^r_n, ~~ \bar N^{r3}_{n0} = 0. 
\eeq 
\end{subequations}
The symmetric properties of the Neumann functions follow from Eq. (\ref{neumandef}): 
\beq
\bar N^{rs}_{nm} = \bar N^{sr}_{mn},~~~ n, m \ge 0. 
\eeq 
In terms of the Neumann functions, the Fock space representation of the three-string vertex is obtained 
as 
\begin{subequations}
\beq
\vert \bbV_{[3]} \rangle &=& (2\pi)^d \d \left( \sum_{r=1}^3 p_r \right) \exp \bigl[(E[1,2;3] \bigr] \vert 0 \rangle, \label{v3fock}\\
E[1,2;3] &=& \half \sum_{n,m =1}^\infty\sum_{r,s=1}^3 \bar N^{rs}_{nm} \a^{(r)}_{-n} \cdot \a^{(s)}_{-m} + 
\sum_{n =1}^\infty \sum_{r=1}^3\bar N^r _n \a^{(r)}_{-n} \cdot \bbP \nn\\
&& +  \t_0\sum_{r=1}^3 \frac{1}{\a_r} \left(\frac{(p^{(r)})^2}{2}-1 \right), \label{E123}
\eeq
where $ 
\a^{(r)}_n = \sqrt{n} a^{(r)}_n, ~~~ \a^{(r)}_{-n}= \sqrt{n} a^{(r)\dag}_n $. 
\end{subequations}
Now we may write the cubic interaction term of the string field theory in the proper-time gauge 
as follows
\beq
{\cal S}_{[3]} &=& \int \prod_{r=1}^3 dp^{(r)} \d \left(\sum_{r=1}^3 p^{(r)}\right) \frac{2g}{3}\, 
\text{tr}\, \langle \Psi^{(1)}, \Psi^{(2)}, \Psi^{(3)} \vert E_{[3]}[1,2,3] \vert 0 \rangle \label{cubicterm}
\eeq 
where $g$ is the string coupling constant. 

\subsection{Three-string vertex in the zero-slope limit}

The next step is to calculate the three-string scattering amplitude in the zero-slope limit 
to confirm that the covariant three-string vertex in the proper-time gauge 
Eq. (\ref{v3fock}) yields the correct 
three-gauge-field vertex of the non-Abelian Yang-MIlls gauge theory. In the zero-slope limit,
the external string state is the massless gauge field vector, which is obtained by
applying $a^{(r)\dag}_1$, $r=1,2,3$ to the string vacuum state
\beq
\vert \Psi^{(1)}, \Psi^{(2)}, \Psi^{(3)} \rangle = \prod_{r=1}^3 A(p^{(r)}) \cdot a_1^{(r)\dag} \vert 0 
\rangle. 
\eeq
Thus, it suffices to evaluate the Neumann functions $\bar N^{rs}_{11}$ and $\bar N^r_1$ only:
\begin{subequations}
\beq
\bar N^{11}_{11} &=& \frac{1}{2^4}, ~~~ \bar N^{22}_{11} = \frac{1}{2^4}, ~~~ \bar N^{33}_{11} = 2^2, \label{neumanna}\\
\bar N^{12}_{11} &=& \bar N^{21}_{11} = \frac{1}{2^4}, ~~~ \bar N^{23}_{11} = \bar N^{32}_{11} = \half, 
~~~ \bar N^{31}_{11} = \bar N^{13}_{11} = \half ,\label{neumannb}\\
\bar N^1_1 &=& \bar N^2_1 = \frac{1}{4}, ~~~ \bar N^3_1 = -1 . \label{neumannc}
\eeq
\end{subequations}

The momentum variable $p^{(r)}$ should be replaced by $\sqrt{\ap} p^{(r)}$ 
if we restore the slope parameter $\a^\prime$. Consequently, in the  
zero-slope limit, we may drop the third term in $E[1,2;3]$, Eq. (\ref{E123})
\beq
\t_0 \sum_{r=1}^3 \frac{1}{\a_r} \frac{p^2_r}{2} \longrightarrow 0 .
\eeq
This observation greatly simplifies our calculation in the zero-slope limit. 
The cubic string field interaction term, given by Eq. (\ref{cubicterm}) may reduce in the zero-slope limit, to the following three-gauge-field term
\beq
S_{\text{Gauge[3]}} &=&\frac{2g_{YM}}{3} e^{- {\t_0} \sum_{r=1}^3 \frac{1}{\a_r} }\int \prod_{i=1}^3 \frac{dp^{(i)}}{(2\pi)^d} (2\pi)^d \d \left(\sum_{i=1}^3 p^{(i)} \right) \nn\\
&& \text{tr} \Bigl\langle 0 \Bigl \vert
\left\{\prod_{i=1}^3 A(p^{(i)}) \cdot a^{(i)}_1 \right\} \half \sum_{r, s =1}^3\bar N^{rs}_{11} \left(a^{(r)\dag}_1 \cdot a^{(s)\dag}_1 \right) \sum_{t=1}^3 \bar N^{t}_1 \left(a^{(t)\dag}_1 \cdot 
\bbP \right)\Bigl \vert 0 \Bigl \rangle \nn\\
&=& \frac{2g_{YM}}{3} \times 2^3 \times \half \times 2 \int \prod_{i=1} dp^{(i)} \d \left(\sum_{i=1}^3 p^{(i)} \right) (p^\m_2 - p^\m_1) \nn\\
&& \text{tr} \Bigl( \bar N^{12}_{11} \bar N^3_1 A(p_1) \cdot A(p_2) A(p_3)_\m + 
\bar N^{23}_{11} \bar N^1_1 A(p_2) \cdot A(p_3) A(p_1)_\m  \nn\\
&& + \bar N^{31}_{11} \bar N^2_1 A(p_3) \cdot A(p_1) A(p_2)_\m 
\Bigr)
\eeq 
where $g_{YM}$ is the Yang-Mills coupling constant, related to the string interaction coupling $g$ as 
\beq
g_{YM} =\left(\ap\right)^{\frac{d}{4}-1} g. 
\eeq 

Making use of the Neumann functions in the proper-time gauge Eqs. (\ref{neumanna}), (\ref{neumannb}) and 
(\ref{neumannc}), we have
\beq
S_{\text{Gauge}[3]} &=& \frac{2g_{YM}}{3} \, 2^3  \int \prod_{i=1} dp^{(i)} \d \left(\sum_{i=1}^3 p^{(i)} \right) (p^\m_2 - p^\m_1) \nn\\
&& \text{tr} \Bigl(-\frac{1}{2^4} A(p_1) \cdot A(p_2) A(p_3)^\m + 
\frac{1}{2^3} A(p_2) \cdot A(p_3) A(p_1)^\m + \frac{1}{2^3} A(p_3) \cdot A(p_1) A(p_2)^\m
\Bigr) \nn\\
&=& \frac{2g_{YM}}{3} \, 2^3 \left(\frac{1}{2^4}+ \frac{1}{2^3} \right)  \int \prod_{i=1} dp^{(i)} \d \left(\sum_{i=1}^3 p^{(i)} \right) (p^\m_1 - p^\m_2) \,\text{tr}\, \Bigl( A(p_1) \cdot A(p_2) A(p_3)^\m \Bigr) \nn\\
&=& g_{YM}  \int \prod_{i=1} dp^{(i)} \d \left(\sum_{i=1}^3 p^{(i)} \right) p^\mu_1 
\,\text{tr} \Bigl( A^\n(p_1)  \left[ A_\n(p_2), A_\m(p_3)\right] \Bigr). 
\eeq 
The reduced three-string interaction term can be identified as the momentum space representation of 
the three-gauge-field interaction term of the $SU(N)$ non-Abelian group Yang-Mills gauge theory 
\beq
S_{\text{Gauge}[3]} 
&=& g_{YM}  \int d^d x ~ i\,\text{tr} \left(\p_\m A_\n - \p_\n A_\m \right) \left[ A^\m, A^\n \right] \label{gauge3}. 
\eeq 
By an explicit calculation, we find that the three-string vertex of the covariant string field theory 
in the proper-time gauge correctly reduces to the three-gauge-field vertex of the non-Abelian gauge 
field theory. Note that the $U(1)$ component of the string field $\Psi^0[X^{(r)}]$ does not enter the 
three-string vertex. It is also consistent with the Abelian gauge field theory, which does not contain 
the three-gauge-field vertex. Because the free string propagator does not depend on the length parameters, 
the free string field action and the covariant three-string vertex may constitute a consistent covariant 
string field theory. It remains to be seen whether the four-gauge-boson vertex of the Yang-Mills 
gauge theory also arises as a part of low energy effective action from the tree string diagram of the covariant string field theory in the proper-time gauge or not.  
If the contact four-gauge-field term of the Yang-Mills theory is not obtained as 
a part of effective action of 
the covariant string field theory with the three-string vertex only, we may have to extend the theory 
by introducing an additional four-string vertex term in the classical action of the open 
string field theory. 

\section{The Effective Quartic String Field Interaction}

\subsection{Effective four-string vertex at tree level}

\begin{figure}[htbp]
   \begin {center}
    \epsfxsize=0.8\hsize

	\epsfbox{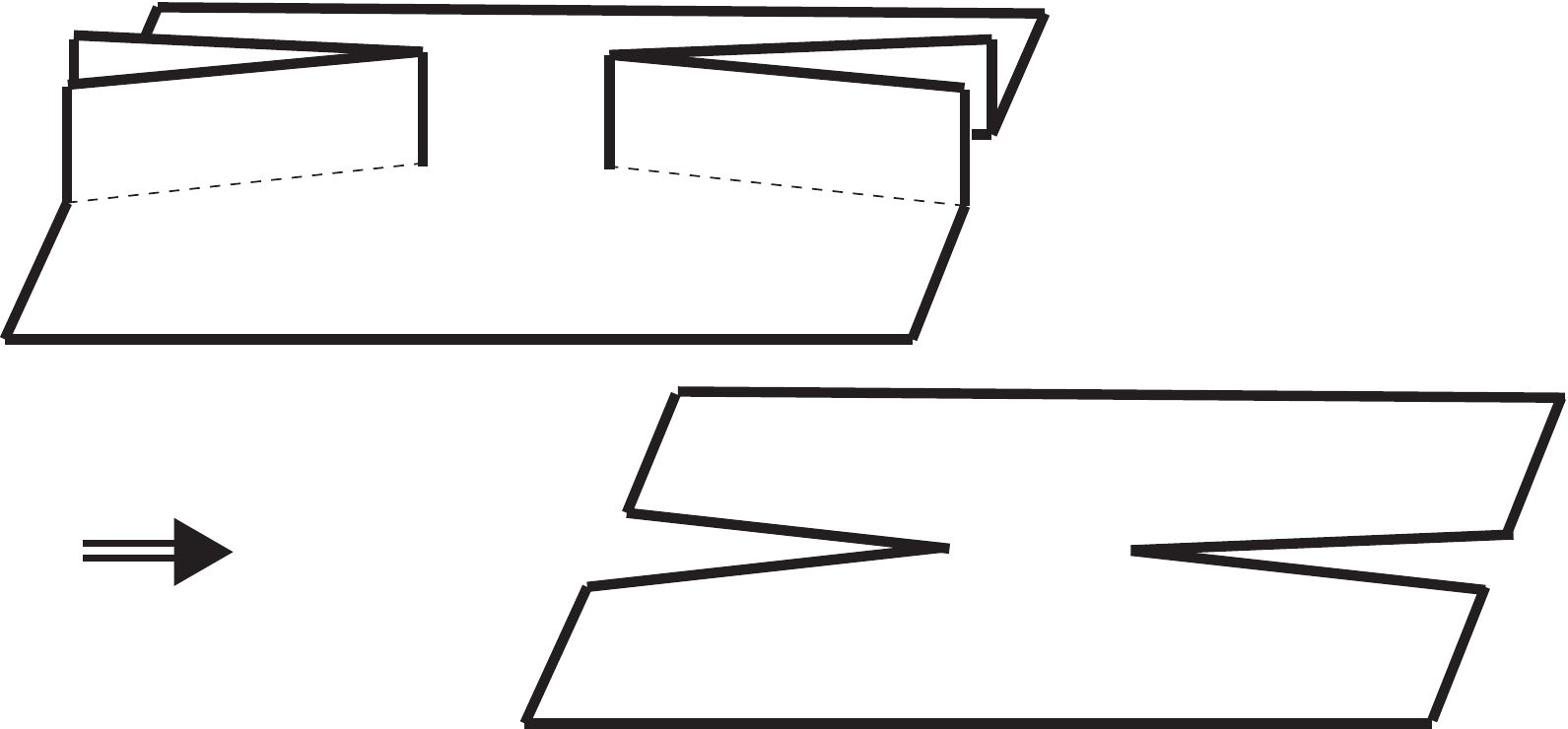}
   \end {center}
   \caption {\label{deformfour} Deformation of the four-string scattering diagram}
\end{figure}

As we define a covariant perturbative string field theory with the three-string vertex, we find that the 
perturbation theory generates at tree level the four-string vertices which consist of two cubic string vertices and a propagator. The purpose of examining those diagrams of four-string vertices is twofold:
1) We may confirm that diagrams of the string perturbation theory correctly reduce to those of the Yang-Mills
gauge theory in the zero-slope limit. 2) We will investigate if the quartic gauge field coupling 
of the Yang-Mills gauge theory may be generated by the tree level diagrams of string field theory with the cubic string vertex in the zero-slope limit. 
We may formally write the four-string scattering amplitude as follows:
\beq
{\cal F}_{\text{Full} [4]} &=& \int D[\Psi]\, \text{tr} \prod_{r=1}^4 \Psi^{(r)}\exp \left[ - \int \text{tr} \left(
\Psi * Q \Psi + \frac{2g}{3} \Psi * \Psi * \Psi \right) \right]. 
\eeq 
(This is in fact the full four-string Green's function.) If we choose the Siegel gauge to fix the 
BRST gauge invariance, $L_0$ may replace the BRST operator $Q$ in the kinetic term. At tree level,
the four-string scattering amplitude is read as 
\beq
{\cal F}_{\text{Tree}[4]} &=& \int D[\Psi]\, \text{tr} \prod_{r=1}^4 \Psi^{(r)} \frac{1}{2!} 
\left(\frac{2g}{3}\right)^2 \left[\int \text{tr} \left(\Psi * \Psi* \Psi \right) \right]^2 
e^{\left[-i \int \text{tr} \Psi L_0 \Psi \right]}\, .
\eeq 
By taking the Wick contraction we may get nine identical diagrams. If we apply the consistent deformation 
to the mid-point overlapping interactions, we may obtain the planar diagram of the string field theory in the
proper time gauge as depicted in Fig. \ref{deformfour} 

\begin{figure}[htbp]
   \begin {center}
    \epsfxsize=0.8\hsize

	\epsfbox{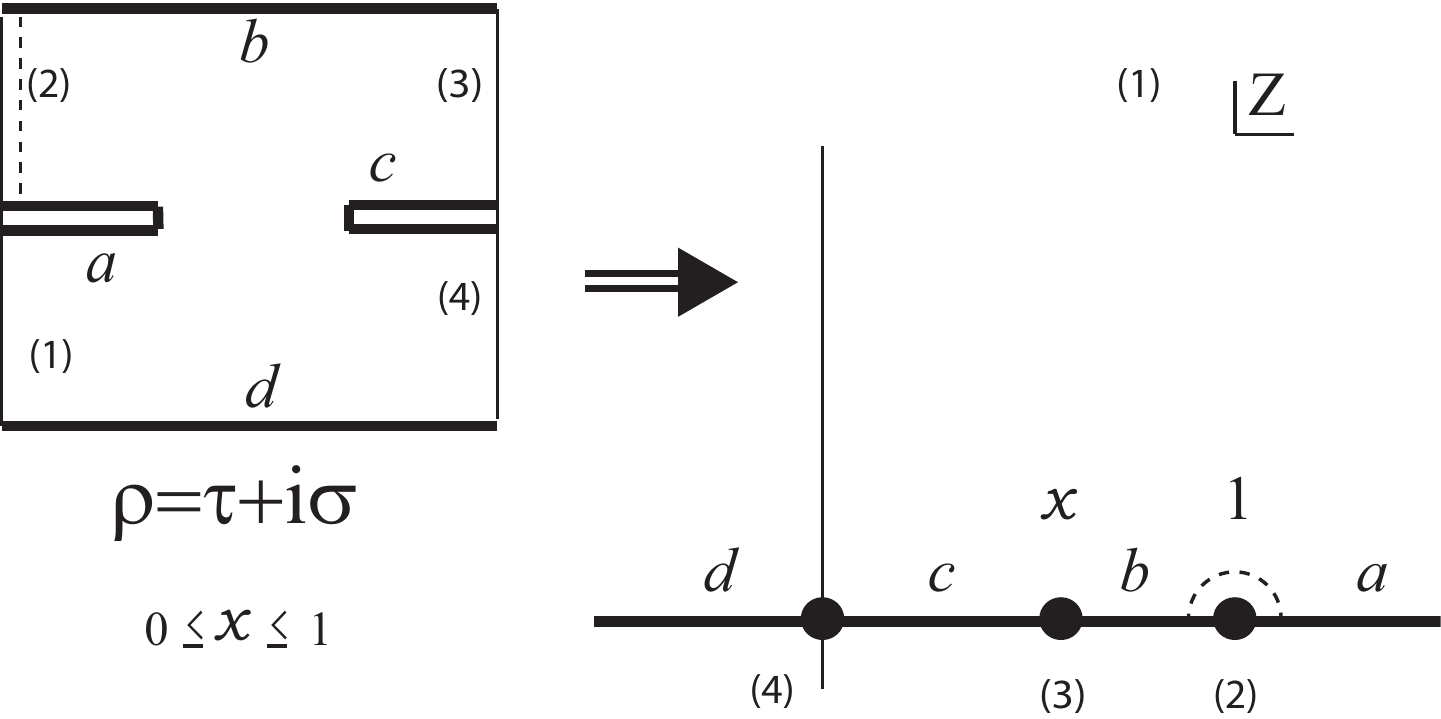}
   \end {center}
   \caption {\label{neumann4} Four-string diagram mapped onto the upper half complex plane.}
\end{figure}

The deformed four-string scattering diagram corresponds to that of planar diagram with length
parameters, 
\beq
\a_1=1, ~~ \a_2 =1, ~~ \a_3 =-1, ~~ \a_4 = -1 .
\eeq
Taking this into account, we may rewrite the four-string scattering amplitude at tree level as 
\beq
{\cal F}_{[4]} &=& 3^2 \times \frac{1}{2!} \left(\frac{2g}{3}\right)^2 \text{tr} \,
\Bigl\langle \Psi^{(1)}, \Psi^{(2)}, \Psi^{(5)} \Bigl\vert \bbV_{[3]}[1,2;5] \Bigr\rangle \nn\\
&&  \Bigl\langle \Psi^{(5)} \Bigl\vert
\frac{1}{L_0-i\e} \Bigr\vert \Psi^{(6)} \Bigr\rangle \Bigl\langle \bbV_{[3]}[3,4;6] \Bigl\vert \Psi^{(3)}, \Psi^{(4)}, \Psi^{(6)} \Bigr\rangle .
\eeq 
We may map the planar four-string scattering diagram onto the upper-half complex plane by using the 
Schwarz-Christoffel transformation as shown in Fig. \ref{neumann4}
\beq
\rho &=& \sum_{r=1}^4 \a_r \ln (z-Z_r)= \ln(z-1) - \ln z - \ln (z-x), \\
Z_1 &=& \infty, ~ Z_2 =1, ~ Z_3 = x, ~ Z_4=0  \nn
\eeq 
where the parameter, $0 \le x \le 1 $ is identified as the Koba-Nielsen variable of the four-string scattering. 
We may recast the four-string scattering amplitude into a $SL(2,R)$ invariant form in term of the 
four-string vertex by using the Cremmer-Gervais identity \cite{Cremmer74}
\begin{subequations} 
\beq
{\cal F}_{[4]} &=& 2g^2 \int \left\vert\frac{\prod_{r=1}^4 dZ_r }{ dV_{abc}}\right\vert 
\prod_{r<s} \vert Z_r - Z_s \vert^{p_r \cdot p_s} \exp\left[-\sum_{r=1}^4 \bar N^{[4]rr}_{00} \right]\nn\\
&& ~~~~~~~~~~~~~~~~~~~~~~~\text{tr} \bigl\langle \Psi^{(1)}, \Psi^{(2)}, \Psi^{(3)}, \Psi^{(4)} \bigl\vert \exp \left[E_{[4]} \right] \bigr\vert 0 \bigr\rangle ,\\
\left[E_{[4]} \right] \bigr\vert 0 \bigr\rangle &=&  \sum_{r, s =1}^4 \left\{ \half \sum_{m, n \ge 0} 
\bar N^{[4] rs}_{mn}\a^{(r)\dag}_{m} \cdot \a^{(s)\dag}_{n} \right\}\bigr\vert 0 \bigr\rangle . 
\eeq
\end{subequations}

\begin{figure}[htbp]
   \begin {center}
    \epsfxsize=0.6\hsize

	\epsfbox{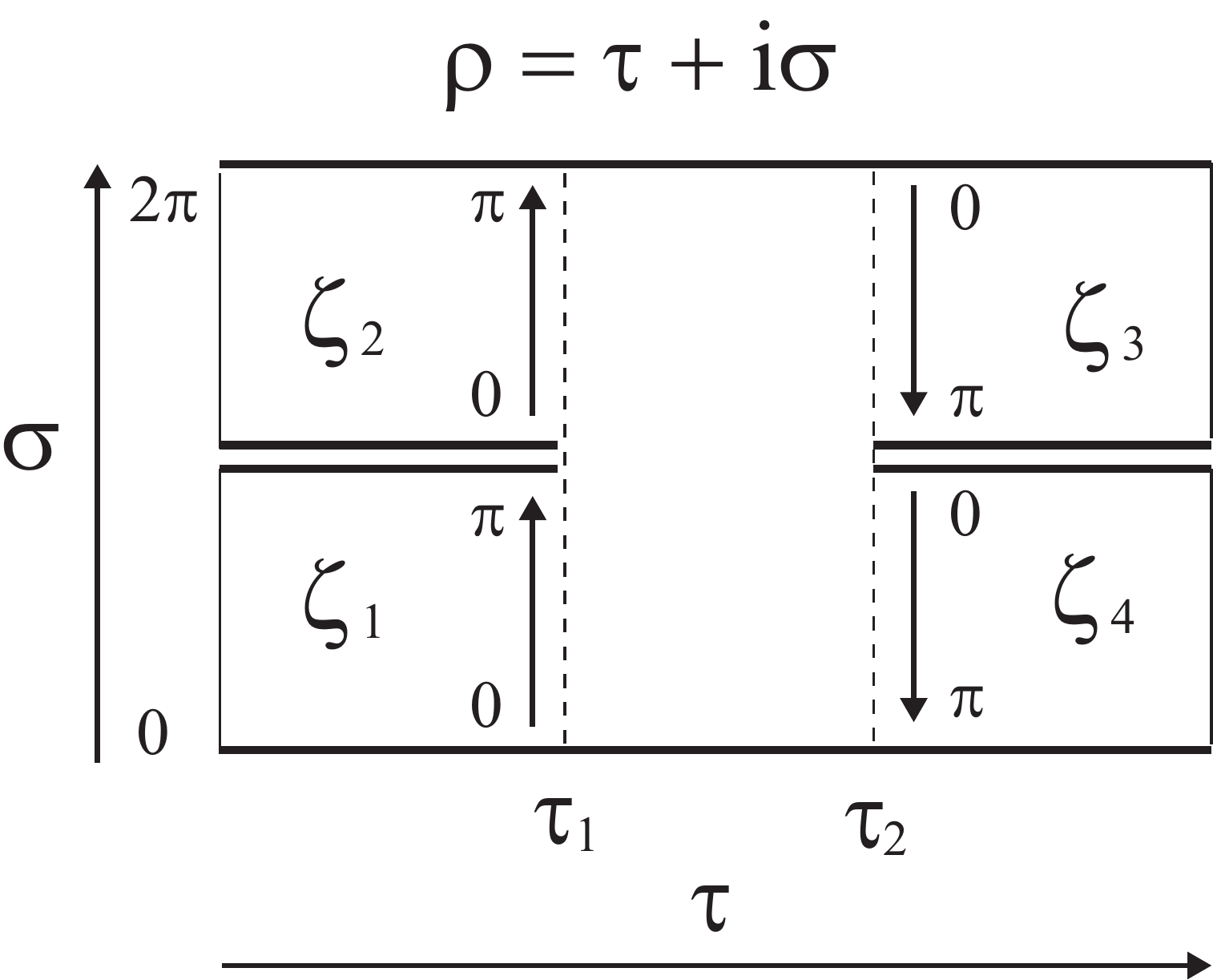}
   \end {center}
   \caption {\label{four} Local coordinates on the four-string scattering diagram}
\end{figure}

The four strings interact at two points on the world sheet, which are determined by the following 
equation 
\beq
\frac{d\rho}{d z} \Bigl\vert = \frac{1}{z-1} - \frac{1}{z-x} - \frac{1}{z} = 0 .
\eeq
It has two real solutions:
\beq
z_\pm = 1 \pm \sqrt{1-x}, ~~~ 0 \le x \le 1 . 
\eeq 
These two solutions define two interaction times on the world sheet of the four-string scattering 
\begin{subequations} 
\beq
\t^{(1)}_0 &=& \t^{(2)}_0 =\t_1 = -2 \ln \left(1+ \sqrt{1-x} \right) < 0, \\
\t^{(3)}_0 &=& \t^{(4)}_0 = \t_2 = -2 \ln \left(1- \sqrt{1-x} \right) > 0 . 
\eeq
\end{subequations}
The local coordinates on individual string patches and the global coordinate on the four-string 
scattering diagram are depicted in Fig. \ref{four}. The local coordinates $\zeta_r$, $r=1,2,3$ may be 
written in terms of $z$ as follows: 
\begin{subequations}
\beq
e^{-\zeta_1} &=& e^{\t_1} \frac{z(z-x)}{1-z}, \\
e^{-\zeta_2} &=& - e^{\t_1} \frac{z(z-x)}{1-z}, \\
e^{-\zeta_3} &=&  e^{-\t_2} \frac{(1-z)}{(z-x) z},\\
e^{-\zeta_4} &=& - e^{-\t_2} \frac{(1-z)}{(z-x)z} .
\eeq
\end{subequations}
The Neumann functions may be computed by using contour integrals on the upper half complex plane \cite{Hata1986,Mandelstam1973,Cremmer74,GreenSW87}
\begin{subequations} 
\beq
\bar N^{rs}_{00} &=& \left\{ 
\begin{array}{ll}
\ln \vert Z_r -Z_s \vert ,  & ~~\mbox{for} ~~r\not=s,\\ 
-\sum_{i(\not=r)} 
\frac{\a_i}{\a_r} \ln \vert Z_r - Z_i \vert + \frac{1}{\a_r} \t^{(r)}_0 , & ~~
\mbox{for} ~~r=s
\end{array} \right.  \label{N00}\\
\bar N^{rs}_{n0} &=& \bar N^{sr}_{0n} = \frac{1}{n} \oint_{Z_r} \frac{d z}{2\pi i} \frac{1}{z-Z_s} e^{-n\zeta_r(z)}, ~~~n \ge 1, \label{Nn0}\\
\bar N^{rs}_{nm} &=& \frac{1}{nm} \oint_{Z_r} \frac{dz}{2\pi i} \oint_{Z_s} \frac{d \zp}{2\pi i} 
\frac{1}{(z-\zp)^2} e^{-n\zeta_r(z) - m \zeta^\prime_s(\zp)}, ~~~ n, m \ge 1  \label{Nnm}
\eeq
\end{subequations}

\subsection{Four-string vertex in the zero-slope limit}

As in the case of the three-string scattering amplitude, we will evaluate the four-string 
scattering amplitude in the zero-slope limit. Choosing the massless vectors as external string 
states, we will compare the four-string scattering amplitude with the four-gauge field 
scattering amplitude of Yang-Mills gauge theory.
In contrast to the light-cone string field theory \cite{Mandelstam1973, Mandelstam1974, Kaku1974a, Kaku1974b, Cremmer74, Cremmer75} and the covariantized light-cone
string field theory \cite{Hata1986}, the Witten's cubic open string field theory does not 
contain a quartic string interaction term in its action. The string field theory in the proper-time gauge
is related to the Witten's cubic string field theory by a consistent deformation, the
string field theory in the proper time gauge also does not possesses a quartic string 
interaction term in its action. 
Thus, the contact quartic gauge interaction of the 
Yang-Mills gauge field should be obtained solely from the effective four-string scattering amplitude
which is generated perturbatively by the cubic string interaction term.  
It is expected that the four-string scattering amplitude in the zero-slope limit would 
comprise contributions of the contact quartic interaction and the effective four-gauge field 
interaction. By explicit calculations we shall identify them separately. 

To convert the scattering amplitude to the effective action we divide the scattering amplitude 
by $2$ because we have to choose one among two three-string vertices to be contracted with the 
initial states $\langle \Psi^{(1)}, \Psi^{(2)} \vert$. Choosing the external four-string states as 
\beq
\bigl\langle \Psi_1, \Psi_2, \Psi_3, \Psi_4 \bigl\vert =\bigl\langle 0 \vert \left(\prod_{i=1}^4 A(p^{(i)}) \cdot a^{(i)}_1 \right)
\eeq
we have 
\beq\label{s4effective}
S_{[4]} &=& \frac{1}{2} \times 2g_{YM}^2 \int \prod_{r=1}^4 dp^{(r)} \d \left(\sum_{r=1}^4 p^{(r)} \right)\int \left\vert\frac{\prod_{r=1}^4 dZ_r }{ dV_{abc}}\right\vert \nn\\
&&
\prod_{r<s} \vert Z_r - Z_s \vert^{p_r \cdot p_s} \exp\left[-\sum_{r=1}^4 \bar N^{[4]rr}_{00} \right]\text{tr} \bigl\langle 0 \vert \left(\prod_{i=1}^4 A(p^{(i)}) \cdot a^{(i)}_1 \right) \nn\\
&&
\frac{1}{2!} \times \frac{1}{2^2} \left\{\sum_{r, s =1}^4
\bar N^{[4] rs}_{11} a^{(r)\dag}_{1} \cdot a^{(s)\dag}_{1} \right\}^2
\bigr\vert 0 \bigr\rangle .
\eeq
If we contract the string oscillator operators, we would get $4! = 24$ terms. Using the symmetry properties 
of the Neumann functions $\bar N^{rs}_{nm} = \bar N^{sr}_{mn}$, we may rewrite them as a sum of 
three distint terms in the following equation:
\beq \label{s4gauge}
S_{[4]} 
&=& g_{YM}^2\int \prod_{r=1}^4 dp^{(r)} \d \left(\sum_{r=1}^4 p^{(r)} \right) \int \left\vert\frac{\prod_{r=1}^4 dZ_r }{ dV_{abc}}\right\vert \prod_{r<s} \vert Z_r - Z_s \vert^{p_r \cdot p_s} \nn\\
&&
\exp\left[-\sum_{r=1}^4 \bar N^{[4]rr}_{00} \right] 
\text{tr}
\Biggl (\bar N^{12}_{11} \bar N^{34}_{11} A^\m(p_1) A_\m(p_2) A^\n(p_3) A_\n(p_4) \nn\\
&& + \bar N^{13}_{11} \bar N^{24}_{11} A^\m(p_1) A^\n(p_2) A_\m(p_3) A_\n(p_4) 
+ \bar N^{14}_{11} \bar N^{23}_{11} A^\m(p_1) A^\n(p_2) A_\n(p_3) A_\m(p_4)
\Biggr) .~~~~
\eeq
The following expressions may be useful when we evaluate $S_{[4]}$ explicitly:  
\begin{subequations}
\beq 
\left\vert\frac{\prod_{r=1}^4 dZ_r }{ dV_{abc}}\right\vert &=& Z_1^2 dx, \label{s4gaugea}\\
\exp\left[-\sum_{r=1}^4 \bar N^{[4]rr}_{00} \right] \bar N^{12}_{11} \bar N^{34}_{11} 
&=& \frac{1}{Z_1^2} \frac{1}{x^2},  \label{s4gaugeb}\\
\exp\left[-\sum_{r=1}^4 \bar N^{[4]rr}_{00} \right] \bar N^{13}_{11} \bar N^{24}_{11} 
&=& \frac{1}{Z_1^2}, \label{s4gaugec}\\
\exp\left[-\sum_{r=1}^4 \bar N^{[4]rr}_{00} \right] \bar N^{14}_{11} \bar N^{23}_{11} 
&=& \frac{1}{Z_1^2} \frac{1}{(1-x)^2} \label{s4gauged} 
\eeq
\end{subequations}
and in the zero-slope limit,
$\prod_{r<s} \vert Z_r - Z_s \vert^{p_r \cdot p_s} = x^{-\frac{s}{2}} (1-x)^{-\frac{t}{2}}$ .
In the zero-slope limit, we may take 
$\left(p^{(r)}\right)^2 \rightarrow 0$, $r=1, 2, 3, 4$, but 
we should keep finite the Mandelstam variables defined as 
\beq
s = -(p_1+p_2)^2, ~~~ t = -(p_1+p_4)^2, ~~~ u = -(p_1+p_3)^2 . 
\eeq
In the zero-slope limit, the effective four-gauge field action may be expressed as 
\beq \label{4gauge}
S_{[4]} &=& g_{YM}^2\int \prod_{r=1}^4 dp^{(r)} \d \left(\sum_{r=1}^4 p^{(r)} \right) \int_0^1 dx\,\, \text{tr} \Bigl(x^{-\frac{s}{2}} (1-x)^{-\frac{t}{2}} A^\m(p_1) A^\n(p_2) A_\m(p_3) A_\n(p_4)  + \nn\\
&&  + 2 x^{-\frac{s}{2}-2} (1-x)^{-\frac{t}{2}}  A(p_1)^\m A(p_2)_\m A(p_3)^\n A(p_4)_\n\Bigr) \nn\\
&=& g_{YM}^2\int \prod_{r=1}^4 dp^{(r)} \d \left(\sum_{r=1}^4 p^{(r)} \right) \, \text{tr} \, 
\Biggl(A^\m(p_1)  A^\n(p_2) A_\m(p_3) A_\n(p_4) \nn\\
&&+ \frac{2u}{s} A^\m(p_1)  A_\m(p_2) A^\n(p_3) A_\n(p_4) \Biggr)
\eeq 
where we make use of 
\beq 
\int^1_0 dx x^{-\frac{s}{2}} (1-x)^{-\frac{t}{2}} = 1, ~~~~
\int^1_0 dx x^{-\frac{s}{2}-2} (1-x)^{-\frac{t}{2}} = \frac{u}{s} 
\eeq
in the zero-slope limit. 
The resultant effective action $S_{[4]}$ may be compared with the effective four-gauge field 
action which contains contributions both from the contact quartic gauge field interaction and the 
effective quartic gauge field interaction, generated perturbatively by the cubic gauge field 
interaction. In order to identify them separately we shall calculate the effective 
quartic gauge field interaction term. 

In the zero-slope limit, we have shown that the three-string interaction term reduces to the 
three-gauge field interaction term 
\beq
S_{\text{Gauge}[3]} 
&=& g_{YM} \int \prod^3_{i=1} dp^{(i)} \d \left(\sum_{i=1}^3 p^{(i)} \right) (p^\m_1 - p^\m_2) \,\text{tr}\, \Bigl( A(p_1) \cdot A(p_2) A(p_3)^\m \Bigr) . 
\eeq
This three-gauge field interaction perturbatively generates the effective four-gauge field interaction,
mediated by the massless gauge field.  
We can easily calculate the effective four-gauge field interaction term by using usual Feynman diagram \cite{Zee} 
\beq
S_{\text{Massless}[4]} &=& \frac{1}{2!} \times(2!) \, g^2_{YM} \int \prod^4_{i=1} dp^{(i)} \d \left(\sum_{i=1}^4 p^{(i)} \right) \, \text{tr}\, \Bigl(A(p^{(1)}) \cdot A(p^{(2)})\left(p^{(1)}_\m- p^{(2)}_\m\right) \nn\\
&& \frac{\eta^{\m\n}}{\left(p^{(1)} + p^{(2)}\right)^2} \left(p^{(4)}_\n- p^{(3)}_\n \right) 
A(p^{(3)}) \cdot A(p^{(4)})
\Bigr).
\eeq 
It may rewritten in the zero-slope limit as 
\beq
S_{\text{Massless}[4]} &=& g_{YM}^2 \int \prod^4_{i=1} dp^{(i)} \d \left(\sum_{i=1}^4 p^{(i)} \right) \,
\left(1 + \frac{2u}{s} \right) \nn\\
&& \text{tr}\, \Bigl(A(p^{(1)}) \cdot A(p^{(2)})A(p^{(3)}) \cdot A(p^{(4)}) \Bigr). 
\eeq
Thus, the effective four-gauge field action Eq. (\ref{4gauge}) may be written as a sum of the contact
quartic gauge field action $S_{\text{Gauge}[4]}$ and the effective four-gauge field interaction 
mediated by the massless gauge field $S_{\text{Massless}[4]}$ \cite{Hata1986}:
\begin{subequations}
\beq
S_{[4]} &=& S_{\text{Gauge}[4]} + S_{\text{Massless}[4]}, \\
S_{\text{Gauge}[4]}&=& g_{YM}^2 \int \prod^4_{i=1} dp^{(i)} \d \left(\sum_{i=1}^4 p^{(i)} \right) 
\,\text{tr} \, \Biggl(A^\m(p_1) A^\n(p_2) A_\m(p_3) A_\n(p_4)\nn\\ 
&&  - A^\m(p^{(1)}) A_\m(p^{(2)}) A^\n(p^{(3)}) A_\n(p^{(4)}) \Biggr) \nn\\
&=& \frac{g_{YM}^2}{2} \int d^d x \, \text{tr} \, \left[A^\m, A^\n \right]\left[A_\m, A_\n \right].
\eeq 
\end{subequations}
Therefore, we may conclude that the open string field theory in the proper time gauge correctly reproduces 
the Yang-Mills gauge field action in the zero-slope limit. This result also confirms that the Witten's 
cubic open string field theory reduces to the Yang-Mills gauge field in the zero-slope limit
if deformed appropriately.  

\section{Interacting Open String in the Proper Time Gauge}

\begin{figure}[htbp]
   \begin {center}
    \epsfxsize=0.8\hsize

	\epsfbox{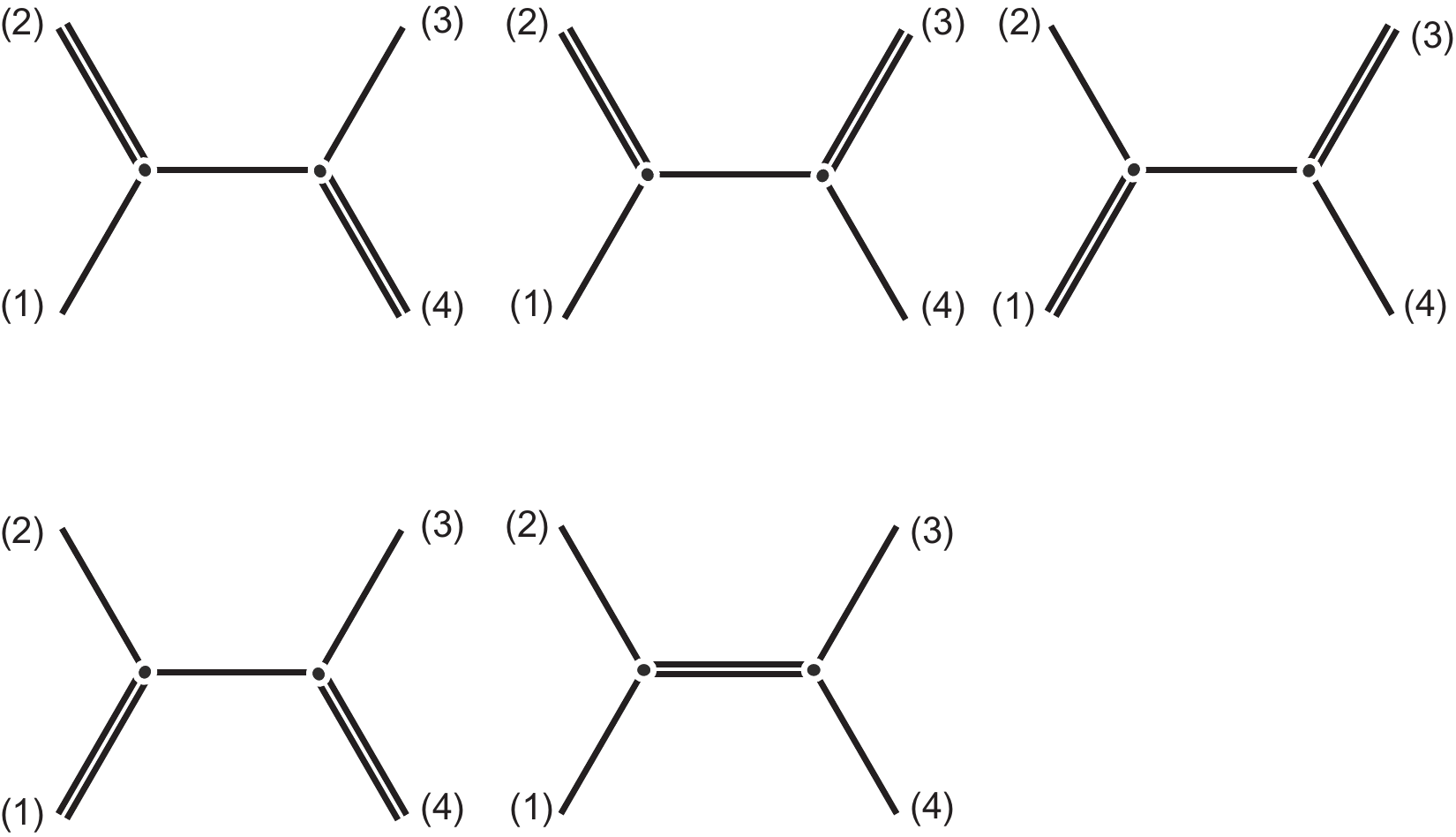}
   \end {center}
   \caption {\label{open5} Effective four-string vertices at tree level}
\end{figure}

We obtained the contact quartic interaction of the gauge field from the string field theory by deforming the world sheet diagram of the four-string scattering. We may deform the cubic open string field 
theory even at the level of action, taking ${\cal S} = {\cal S}_0 + {\cal S}_{[3]}$ as the action for the open string 
field theory:
\beq
{\cal S} &=& \int d^d p \,\,\text{\tr} \, \langle \Psi \vert (L_0 - i\e) \vert \Psi \rangle +
\int \prod_{r=1}^3 dp^{(r)} \d \left(\sum_{r=1}^3 p^{(r)}\right) \frac{2g}{3}\, 
\text{tr}\, \langle \Psi, \Psi, \Psi \vert E_{[3]} \vert 0 \rangle. \label{properaction}
\eeq 
Note that we define the action in Fock space. If the action is defined in the configuration space, 
the string scattering diagrams may not match those of the Yang-Mills gauge field theory in the 
zero-slope limit. This is because that we cannot define the Wick contraction in the configuration space
between two string field operators with different length parameters. If we require that the two 
string field operators to be contracted, have the same length parameter, we would get only five 
tree diagrams by contracting two cubic string vertices as shown in Fig. \ref{open5}  .
The single lines denote the string propagators with length parameter $\vert \a \vert = 1$ and the 
double lines denote the string propagators with the length parameters $\vert \a \vert = 2$.  

\begin{figure}[htbp]
   \begin {center}
    \epsfxsize=0.7\hsize

	\epsfbox{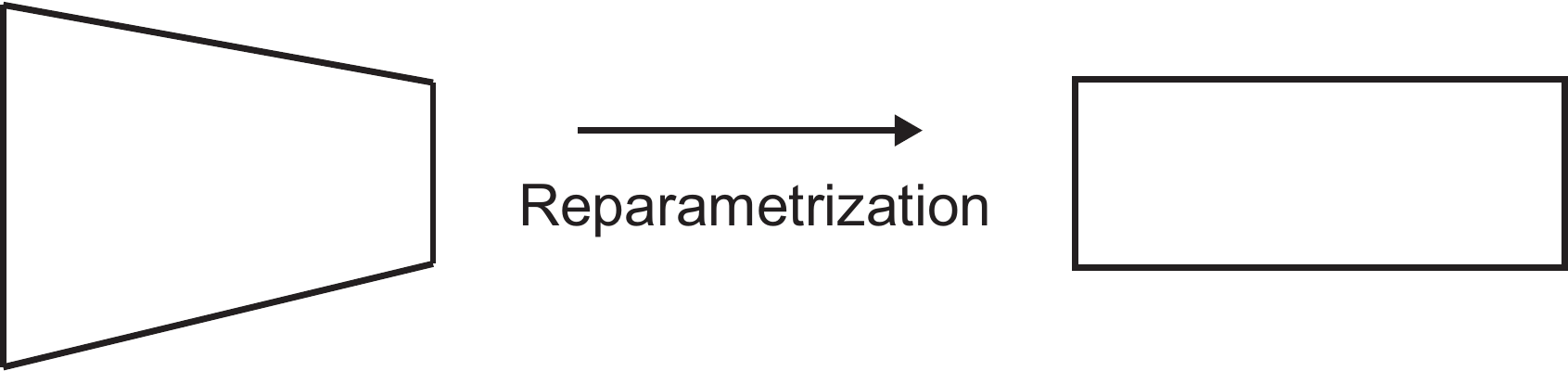}
   \end {center}
   \caption {\label{open8} Wick contraction of two string field operators with different length parameters}
\end{figure}

\begin{figure}[htbp]
   \begin {center}
    \epsfxsize=0.7\hsize

	\epsfbox{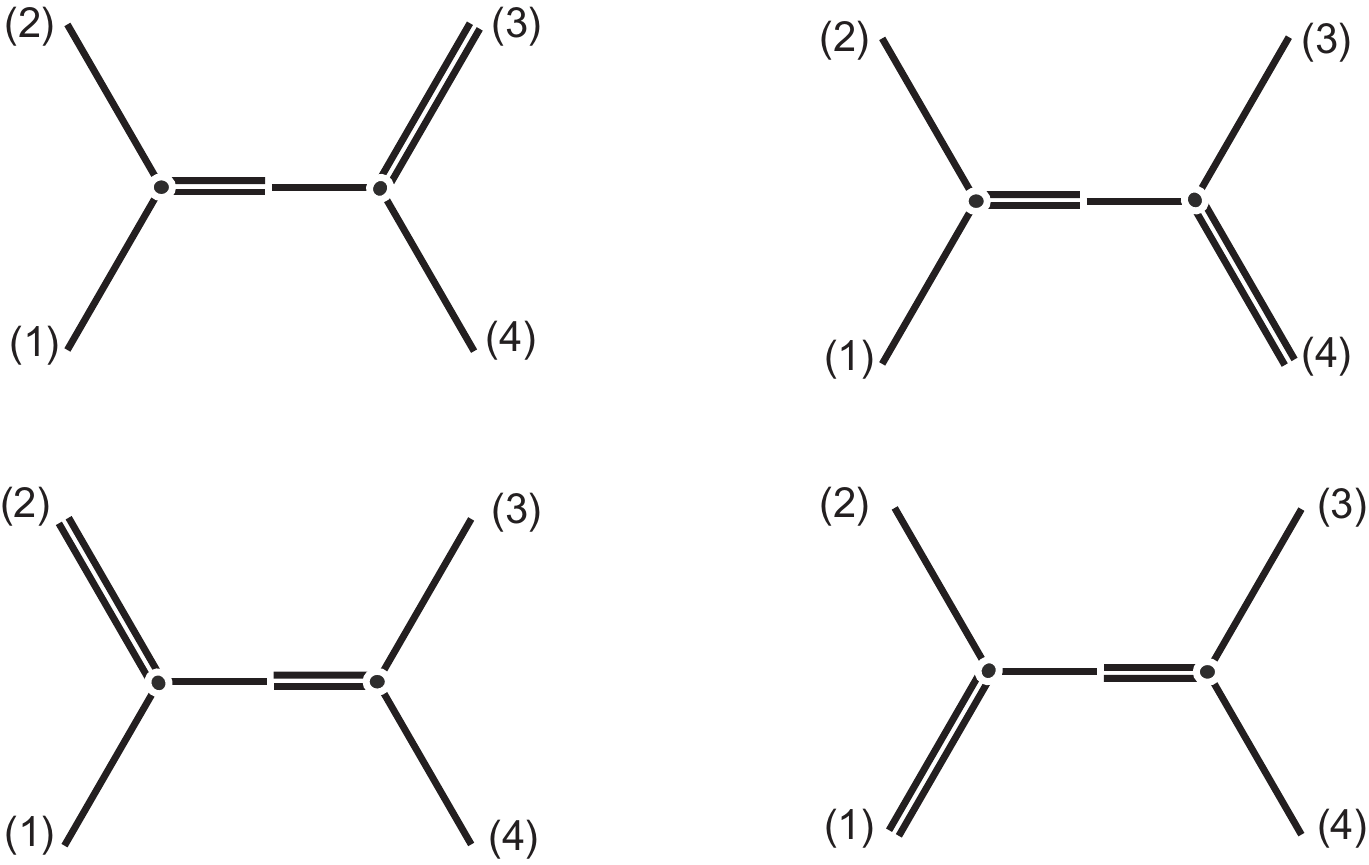}
   \end {center}
   \caption {\label{open9} Additional effective four-string vertices}
\end{figure}

It may be difficult to understand the Wick contraction of a pair of two 
string field operators, which have different length parameters in the string configuration space
$\vert X \rangle$, because the string coordinates with different length parameters have 
different normal mode expansions. However, the length parameter in the proper-time gauge is not a conserved quantity unlike that in the light-cone gauge string field theory. 
The string world sheet corresponding to the Wick contraction of string 
fields with different length parameters 
may be described by the patch of which shape is a trapezoid. It can be 
transformed into a patch of rectangular shape by a two-dimensional reparameterization. (See Fig. \ref{open8}.)
The Polyakov path-integrals over both patches may represent the same free string propagator.  
Hence, we can take the Wick contraction of string fields with different length parameters
with help of the reparameterization on the string world sheet. 
It becomes clear if we write the string state in the number space (Fock space) $\vert \bolk^{(r)}, p^{(r)} \rangle$: Employing the representation of string field operators 
in the number space (Fock space) instead of the conventional 
string configuration space, enables us to define the Wick contraction of string field operators with
different length parameters. 
If we write the string propagator, once in the number space, we find that the string propagator does not depend explicitly on the length parameters of the strings at the edges of the world sheet patch 
\beq
G[(r),(s)] = \Bigl\langle  \bolk^{(r)}, p^{(r)} \Bigl\vert \frac{1}{L_0 - i\e} \Bigr\vert  \bolk^{(s)}, p^{(s)} \Bigr\rangle .
\eeq
In concomitance with it, the kinetic term in the string field action 
$\langle \Psi \vert (L_0 - i\e) \vert \Psi \rangle $ does not depend on the 
length parameter explicitly. The Wick contractions of 
string fields with different length parameters produce additional four-string diagrams
as depicted in Fig. \ref{open9}. As we take the Wick contraction of two cubic string vertices 
in the Fock space, we get nine scattering diagrams, which would match the nine Feynman diagrams of 
Yang-Mills gauge theory in the zero-slope limit. 

\begin{figure}[htbp]
   \begin {center}
    \epsfxsize=0.7\hsize

	\epsfbox{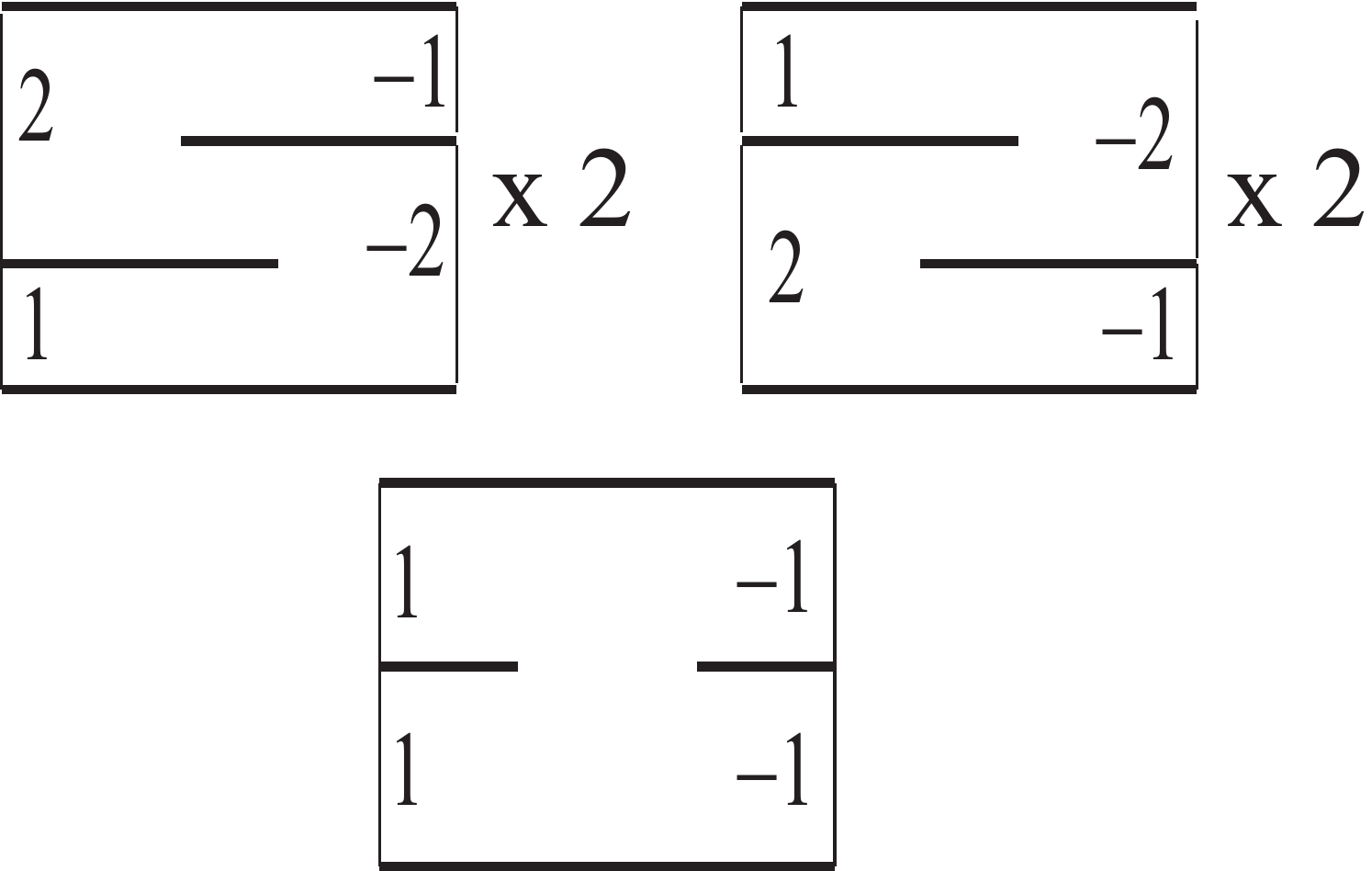}
   \end {center}
   \caption {\label{effective} Planar diagrams of the effective four-string vertices}
\end{figure}

\begin{figure}[htbp]
   \begin {center}
    \epsfxsize=0.5\hsize

	\epsfbox{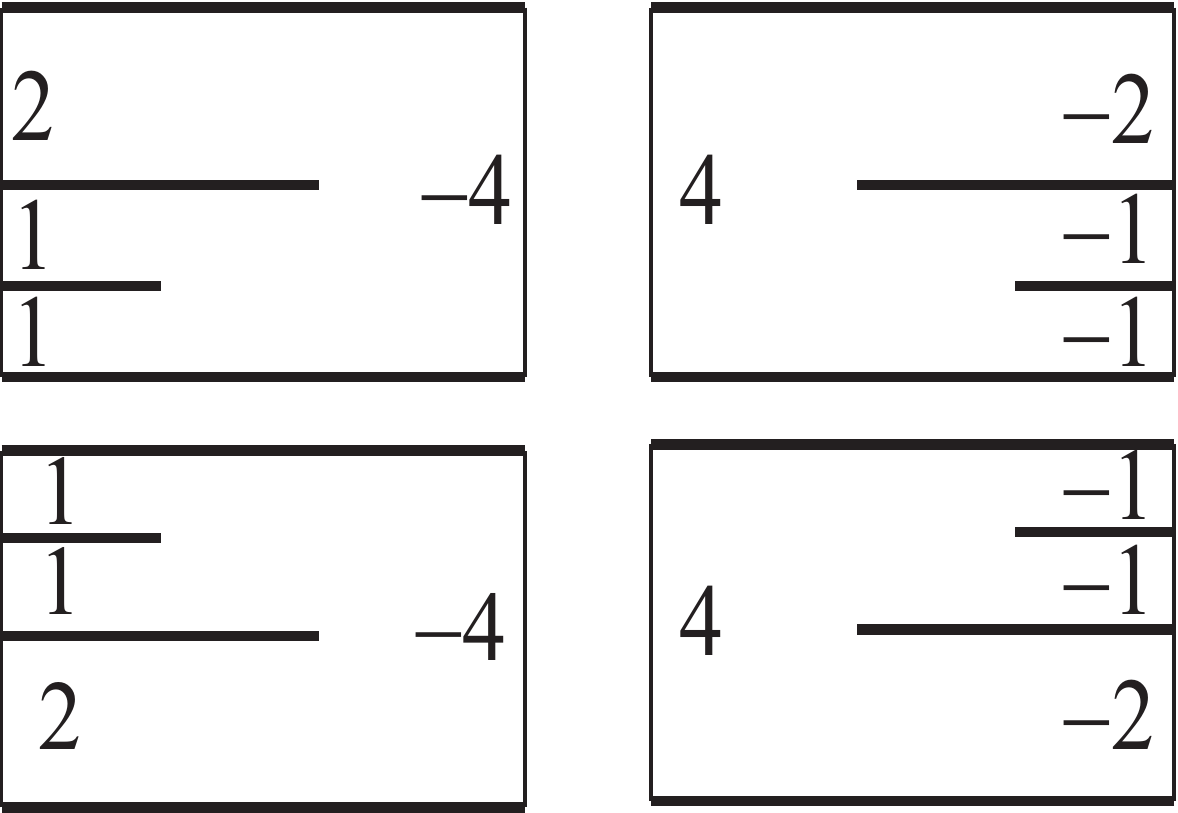}
   \end {center}
   \caption {\label{effective2} Planar diagrams of the additional effective four-string vertices}
\end{figure}

Evaluation of the four-string scattering amplitudes, represented by the nine diagrams, becomes
a manageable task if we covert them into nine planar diagrams. The five diagrams of Fig. \ref{open5} may be 
mapped onto the five planar diagrams of Fig. \ref{effective}. Likewise, the four diagrams of Fig. \ref{open9} may be also mapped onto the corresponding four planar diagrams as depicted by Fig. \ref{effective2}. It follows from 
the equivalence of the three-string vertices: If the ratios between length parameters are same, two three-string vertices are equivalent to each other. This is because the Neumann functions of the cubic string vertices 
depend only on the ratios between the length parameters.


If once the diagrams of four-string scatterings amplitudes are mapped onto the planar diagrams, 
it would not be a difficult task to calculate the four-string scattering amplitude in the zero-slope
limit with massless gauge vectors as external string states. All nine diagrams may be mapped onto 
the upper half complex plane by using the Schwarz-Christoffel transformations with length parameters 
appropriately chosen: 
\begin{subequations}
\beq
\rho &=& \sum_{r=1}^4 \a_r \ln (z- Z_r), \\
Z_1 &=& \infty, ~~Z_2= 1, ~~ Z_3 =x, ~~ Z_4 =0 .
\eeq 
\end{subequations} 
Each diagram contributes to the effective four-gauge field interaction term, which takes the same form as Eq. (\ref{s4effective}), but with different Neumann functions. However, in the zero-slope limit, their contributions to the 
effective four-gauge field interaction become identical. This is due to the fact that the coefficients 
of the effective four-gauge interaction terms do not depend on the length parameters as we can see in Eqs. (\ref{s4gauge}, \ref{s4gaugea} \ref{s4gaugeb}, \ref{s4gaugec}, \ref{s4gauged}). 
Therefore, we would get the same effective four-gauge field action $S_{[4]}$ as Eq. (\ref{4gauge}). 
In the zero-slope limit, the string field action in the proper time gauge Eq. (\ref{properaction}) 
reduces to the 
gauge field action which consists of the kinetic term and the cubic interaction term of the Yang-Mills 
gauge field theory:
\beq
{\cal S} \rightarrow S &=& \int d^d p \, \text{tr} \, \frac{p^2}{2}\, A(p) \cdot A(-p) \nn\\
&& + g_{YM}  \int \prod_{i=1} dp^{(i)} \d \left(\sum_{i=1}^3 p^{(i)} \right) p^\mu_1 
\,\text{tr} \Bigl( A^\n(p_1)  \left[ A_\n(p_2), A_\m(p_3)\right] \Bigr) .
\eeq  
This gauge field action generates perturbatively the same effective four-gauge field interaction which 
is mediated by the massless gauge field, $S_{\text{Massless}[4]}$ as before. 
In the light of the above consideration, we may conclude that the open string field theory action 
in the proper time gauge, which may be obtained by deforming the Witten's cubic open string field theory 
at the level of action, also consistently yields the Yang-Mills gauge field action in the zero-slope limit.

\section{Conclusions}

The covariant second quantized string theories have been studied for more than three decades. However, the previous studies on the covariant string field theories are mostly limited to the free string theories or 
those on a space filling single D-brane. In the present, work we construct a covariant interacting open boson string field theory on multiple D-branes by choosing the proper-time gauge fixing condition. 
In the proper-time gauge, the lapse and shift functions, which parameterize the two-dimensional world sheet, 
are chosen to be zero except for the zero-mode of the lapse function. The BRST ghost structure 
results from the gauge fixing by using the proper-time gauge. The proper-time emerges 
as a product of the time-like coordinate of the world sheet, $\t$ and the constant lapse function, which sets 
the scales of temporal and spatial directions on the world sheet. Evaluating the path integral over a 
strip of the string world sheet in the proper-time gauge, yields the covariant free string field propagator. 

On the multiple D-branes, the string fields take values in the Lie-algebra of $U(N)$ group.  
When we construct the three-string vertex, we should take the $U(N)$ group structure into account carefully. 
In the proper-time gauge, the junction of three strings describes the process through which two identical 
strings join together to form a single string. If we require that the energy scale would be same at the 
junction of the three strings, we may fix the length parameters of the covariant string field theory. 
The resultant three-string vertex can be made equivalent to that of the Witten's cubic string field theory,
by extending the ranges of two strings at initial time, to be joined and replacing the 
end-point interaction by the mid-point overlapping interaction. The extra patches, introduced during this process, would not contribute to the scattering amplitude if the initial string states are appropriately 
chosen. Due to this equivalence, the string field theory in the proper time gauge is 
invariant under the BRST gauge transformation even if the length parameters are completely fixed. 
The Fock space representation of the three-string vertex is obtained by making use of the well-known 
light-cone string field theory technique: This was possible because the world sheet diagram of the string field theory in the proper time gauge is planar. In the zero-slope limit, 
the three-string vertex reduces to the three-gauge field interaction with the correct Yang-Mills 
coupling. Although the three-string vertex in the proper-time gauge appears to be asymmetric, the three-string 
scattering amplitude turns out to be symmetric as we integrate all momentum variables. 
 
Having constructed the covariant three-string 
vertex on the multiple D-branes, we examined the effective four-string vertices which are obtained by 
taking the Wick contraction of two three-string vertices. Expanding the full four-string Green's function
(the four-string scattering amplitude) in string coupling constant $g$ perturbatively
by using the Witten's cubic open string field theory, we have the four-string scattering amplitude at tree level. Deforming the diagram of the four-string scattering as in the case of three-string 
scattering, we transform the non-planar diagram into a planar diagram with length parameters fixed. 
The Fock space representation of the four-string vertex has been constructed by mapping the planar diagram onto the upper half complex plane. In the zero-slope limit, we calculated the four-gauge interaction term
by choosing the four-gauge field vectors as external string states. It is confirmed that the four-gauge 
interaction is precisely the sum of the contact quartic interaction and the effective four-gauge
field interaction, generated perturbatively by the cubic gauge interaction. Thus, it proves that the 
open string field theory on multiple D-branes in the proper time gauge is equivalent to the 
non-Abelian Yang Mills gauge theory in the zero-slope limit. We may deform the cubic open string field theory
even at the level of action by replacing the mid-point overlapping interaction by the 
end-point interaction of the string field theory in the proper time gauge. If the action is defined in the 
Fock space, we would have a consistent perturbation theory, which also reproduces the Yang-Mills gauge 
field action in the zero-slope limit. 

The open string field theory in the proper time gauge or the deformed cubic string field theory may
serve as a practical tool to study particle physics in the framework of string theory. The tachyon condensation \cite{TLee2001,sen1,senreview,coletti,TLee206,forini} and scattering amplitudes with 
scalars particles and vector gauge particles \cite{JCLee2015,Lai2016}
would be immediate applications of the string field theory in the 
proper time gauge. It may be also interesting to study leading stringy corrections \cite{Huang2016a,Huang2016b} to the various particle scattering amplitudes by adopting the string field theory proposed in the present work. 
There is a lot of room to improve the present work: The ghost sector has been suppressed for the sake of simplicity. It may be not difficult to include the ghost sector so that the covariant interacting string theory on multiple D-branes becomes manifestly BRST invariant. It would be possible to construct 
the covariant interacting super-string field theory actions \cite{Green1983a,Green1983b,Green1984,Mandelstam1986} on multiple D-branes along the direction of the present work.


\vskip 1cm

\noindent{\bf Acknowledgments}

This work was supported by 2017 Research Grant from Kangwon National University. 
Part of this work was done during author's visit to IBS (Korea) and NCTU (Taiwan).
He would llike to thank Soo-Jong Rey, Jen-Chi Lee and Yi Yang for their hospitality. 


%





\end{document}